\documentclass[12pt]{iopart}
\usepackage{iopams}
\usepackage{graphicx}
\usepackage{setstack}

\begin{document}




\title{Dynamics of an impact oscillator near a
  degenerate graze}

\author{D R J Chillingworth}

\address{School of Mathematics,
	 University of Southampton,	
         Southampton  SO17~1BJ, UK}

\ead{drjc@maths.soton.ac.uk}


\begin{abstract}
We give a complete analysis of low-velocity dynamics close to grazing
for a generic one degree of freedom impact oscillator. 
This includes nondegenerate (quadratic) grazing and minimally
degenerate (cubic) grazing, corresponding respectively to nondegenerate
and degenerate {\em chatter}. We also describe the dynamics associated
with generic one-parameter bifurcation at a more degenerate (quartic)
graze, showing in particular how this gives rise to the often-observed
highly convoluted structure in the stable manifolds of chattering orbits.
The approach adopted is geometric, using methods from singularity theory.
\end{abstract}


\submitto{Nonlinearity}


\def\ssk{\smallskip}
\def\msk{\medskip}
\def\bsk{\bigskip}
\def\noi{\noindent}
\def\R{{\mathbf R}}
\def\S{{\mathcal S}}
\def\T{{\mathcal T}}
\def\br{{\mathbf R}}
\def\bn{{\mathbf N}}
\def\bz{{\mathbf Z}}
\def\bq{{\mathbf Q}}
\def\eps{\varepsilon}
\def\lam{\lambda}
\def\Lam{\Lambda}
\def\sig{\sigma}
\def\om{\omega}
\def\Om{\Omega}
\def\dyn{ G_c}
\def\sic{\Sigma_c}
\def\vtt{v,\tau;t}
\def\pp{\partial}
\def\grad{\mathop{\mathrm{grad}}\nolimits}
\def\dbd#1#2{\frac{\partial{#1}}{\partial{#2}}}
\def\act{a_c(\tau)}
\def\ppb{\Pi_0^+}
\def\pnb{\Pi_0^-}

\def\proof{\emph{Proof.\enspace}}
\def\qed{\hfill$\Box$}
\def\endproof{\hfill\qed}

\def\nhd{neighbourhood}
\def\til{\tilde}

\def\tfrac#1#2{{\textstyle\frac#1#2}}

\newtheorem{lem}{Lemma}[section]
\newtheorem{prop}[lem]{Proposition}
\newtheorem{cor}[lem]{Corollary}
\newtheorem{theo}[lem]{Theorem}
\newtheorem{conj}[lem]{Conjecture}
\newtheorem{defn}[lem]{Definition}

\section{Introduction}
The theory of forced nonlinear oscillations has well known and
widespread applications
in science and engineering: see Hayashi~\cite{HA}, Fidlin~\cite{FI},
for example, as well as landmark texts Andronov {\em et al.}~\cite{AVK},
Schmidt {\em et al.}~\cite{ST}, Guckenheimer {\em et al.}~\cite{GH}.  
However, there are many engineering contexts in which 
oscillations are a highly undesirable feature of performance, especially
when these lead to impacts between moving parts 
or against rigid supports, generating noise and wear and
ultimately leading to mechanical breakdown: for some examples
see Alzate {\em et al.}~\cite{ABMS}, Mason {\em et al.}~\cite{MHW},
Stewart~\cite{STE}, Pavlovskaia {\em et al.}~\cite{WP} as well as
those described in
books such as Babitsky {\em et al.}~\cite{BA}, Brogliato~\cite{BR} and 
di Bernardo {\em et al.}~\cite{BBCK}.

\medskip

A simple model for such dynamical impact phenomena 
(often called  {\em vibro-impact} systems)
is the one degree of freedom {\em impact oscillator}.  This consists
of a forced nonlinear oscillator 
\begin{equation}  \label{e:eqn}
\ddot x + f(x,\dot x) = g(t),\quad  g(t+T)=g(t)
\end{equation}
for $x\in\R$ and fixed $T>0$, together with a {\em constraint} $x\ge c$ where
$c$ is a constant often called the \emph{clearance}.  
Here $f$ and $g$ are 
smooth\footnote{For simplicity we take {\em
    smooth} to mean $C^\infty$, although many of the results will
  clearly hold with less differentiability.} 
functions, and as usual the dot denotes $d/{dt}$.
\msk

The model must include a prescription for how the
dynamics are to proceed when $x=c$. This will typically be given by a {\em
restitution rule} that whenever $(x,\dot x)=(c,v)$ then the velocity $v$
is instantaneously replaced by $\til v$ (a function of $v$ and
perhaps other variables) satisfying $\til vv<0$.  
There are many variations on this model, involving more than one
degree of freedom and/or more than one constraint or a
non-instantaneous restitution rule.
\msk

The fact that a simple impact system can give rise to extremely
complicated dynamics has long been noted, and there are many
analytical and numerical studies showing complicated periodicities
or convoluted orbit structures in phase space: see for
example 
Thompson {\em et al.}~\cite{TG82},
Shaw {\em et al.}~\cite{SH83}, 
Hindmarsh {\em et al.}~\cite{HI}, 
Peterka {\em et al.}~\cite{PV}, 
Kleczka {\em et al.}~\cite{KKS}, 
Virgin {\em et al.}~\cite{VB}, 
Wagg {\em et al.}~\cite{WB01},\cite{WB04},
Valente {\em et al.}~\cite{VMM}. 
Nevertheless, even a $1$-degree of freedom linear oscillator with
a simple restitution rule is still poorly understood from the overall 
dynamical point of view. 
\msk

A more systematic attack on the global problem was undertaken
by Whiston in a pioneering series of papers~\cite{GW87a},\cite{GW87b},\cite{GW92}; 
this enterprise inspired much subsequent work including Budd {\em et al.}~\cite{BDC},
Lamba~\cite{LAM}, Nordmark~\cite{aN91} as well as the present paper.  
\msk

A key feature of impact oscillator dynamics that gives it a different
flavour from (smooth) nonlinear dynamics is the phenomenon of {\it
  grazing} where a periodic orbit reaches the obstacle with zero
velocity and nonzero acceleration: small perturbations of such orbits
may or may not exhibit impacts. The resulting bifurcation phenomena 
have been studied by many authors including
Budd {\em et al.}~\cite{BD},
Chin {\em et al.}~\cite{CONG94},
Dankowitz {\em et al.}~\cite{HD},
Foale {\em et al.}~\cite{FB92},
Ivanov~\cite{apI94},
Nordmark~\cite{aN91},\cite{aN97}, 
Szalai {\em et al.}~\cite{SO}, Molenaar {\em et al.}~\cite{MWW},
Zhao {\em et al.}~\cite{ZD}.
\msk

Another important aspect of impact oscillator dynamics that attracts particular
interest is the low-velocity behaviour close to impact, characterised
by the phenomenon of {\em chatter}.  On leaving $x=c$
with small positive velocity but negative acceleration
the orbit (i.e. the solution to~(\ref{e:eqn}))
may return to hit $x=c$ again after a short time; after restitution at
the impact the process will repeat, and under suitable conditions
the orbit may come to rest at $x=c$ after a finite time but an infinite
number of such impacts.  It will remain stuck at the obstacle $x=c$ until
such time as the acceleration (from the forcing) becomes positive and it
once again lifts off into the region $x>c$. See for example Demeio {\em et al.}~\cite{DL}
for a simple mechanical realisation of this phenomenon,
or Stone {\em et al.}~\cite{SA}, Valente {\em et al.}~\cite{VMM} for more
complicated dynamical settings.
The first systematic
analytic and geometrical analysis of the dynamics of
chatter was carried out by Budd {\em et al.}~\cite{BD}, using a combination of 
insightful numerics supported theoretically by
first and second order approximations. 
In this paper we extend those results by providing a complete and
rigorous analysis of chatter for generic impact oscilaltor systems~(\ref{e:eqn})
and for generic $1$-parameter families of such systems.

\msk  

A standard technique for studying a forced oscillator such as (\ref{e:eqn})
without the impacts is to focus on the the discrete dynamical system
$F$ that samples the solutions at times $\{nT:n\in\bz\}$: then
periodic solutions of (\ref{e:eqn}) correspond to periodic orbits of $F$, and
so on.  When impacts are introduced, however, this is no longer
appropriate since the influence of the obstacle  on a given trajectory
cannot be expected to be $T$-periodic. A common approach
to dealing with this problem, used by Shaw {\em et al.}~\cite{SH83}, 
Thompson {\em et al.}~\cite{TG82}, Whiston~\cite{GW87b} and others, is first
to regard~(\ref{e:eqn}) as an autonomous system in $\R^3$ by including
the equation $\dot t=1$ and then to
consider the plane $x=c$ and the discrete dynamical system
defined on it by the {\em first return} of trajectories after impacting
the obstacle.  This fits naturally with the geometry of the physical
description of (\ref{e:eqn}), but the resulting discrete system now has
loci of discontinuity that arise from grazing trajectories. It is not
easy in this setting to see how the discontinuities and the associated
dynamical phenomena respond to change in initial conditions and/or
governing parameters such as the clearance.
\msk


In this paper we use an alternative geometric approach, first set out 
in~\cite{DC}, which allows greater insight into the
geometry of discontinuities and their influence on dynamics by
replacing the
picture of curved orbits intersecting a flat obstacle with an equivalent
picture of straightened-out orbits intersecting a curved obstacle:
the {\em impact surface}.
See Section~\ref{s:surf} below for the precise description.  From this
point of view, the problem of analysing the geometry of orbits at and
near grazing becomes more readily amenable to the tools of local
differential topology, dynamical systems and  singularity theory.
\bsk

Taking this approach, we look again at the phenomena of 
\emph{grazing} and \emph{chatter} in the light of the geometry
of the impact surface. We first use this to 
provide rigorous underpinning for the key results
in~\cite{BD} that are based on approximations:
by placing the results in a wider context
we are able to give a full analysis of the local
dynamical phenomena. We then extend this to cover the case of the
discontinuous bifurcations that arise in the \nhd\ of a
doubly-degenerate graze as the clearance or some other parameter is
varied and the degeneracy is unfolded. In particular we exhibit the
key mechanism for the creation of multiple-loop structures in stable
manifolds of chatter points, which is a first step in understanding
the multiple-loop patterns observed numerically for attractors and
basins of attraction of periodic orbits in impact oscillators: 
see~\cite[Figure 17]{GW87b},
\cite[Figures 4 and 19]{BD},
as well as~\cite[Figure~3]{BBCKNTP} and 
\cite{BBCK} (in particular the front cover). 
Taken together, the results of the present paper provide a complete
description of generic low-velocity impact dynamics
and $1$-parameter bifurcations.

\section{Geometrical model for dynamics: the impact surface}  \label{s:surf}
We begin by  recalling the construction described in
Chillingworth~\cite{DC} for modelling
the overall dynamics of a one degree of freedom impact oscillator.
The model is based on a $2$-dimensional surface in three dimensions
that characterizes the impacts, together with certain natural maps
associated with it that represent the dynamics.
A similar geometrical construction was used by Sotomayor et
al.~\cite{SOT}, albeit with a different dynamical interpretation, to classify
local vector fields on a manifold with boundary. 

\subsection{The impact surface and associated maps}

Let $x(b,v,\tau;t)$ denote the unique solution to~(\ref{e:eqn}) with
initial data 
$$(x,\dot x)=(b,v) \quad \mbox{when}\quad t=\tau.$$  
For fixed choice of $c\in\br$ we then write
\[
x_c(\tau,v;t):=x(c,v,\tau;\tau+t)
\]
so that
\begin{equation}
x_c(\tau,v;0)=c,\quad \dot x_c(\tau,v;0)=v.
\end{equation}
\begin{defn}
The \emph{impact surface} $V_c$ is where $x_c-c\,$ vanishes, that is
\[
V_c:=\{(\tau,v;t)\in\br^3:x_c(\tau,v;t)=c \}. 
\]
\end{defn}
Note that in~\cite{DC} the convention was to write $(v,\tau;t)$ rather than
$(\tau,v;t\,)$ : we trust no confusion will arise from this notation reversal.
\medskip

By definition $V_c$ contains the plane $t=0$, which we shall denote by
$\Pi$ and which plays a key role in our formulation of the dynamics.
  It is proved in~\cite[Lemma 1]{DC} that $V_c$ is indeed a surface
(smooth $2$-manifold), with the exception of the points on the $\tau$-axis
$\Pi_0$ and possibly certain other points where $v=0$, characterised 
precisely in terms of the functions $f$ and $g$ in~(\ref{e:eqn}) and
the clearance $c\,$. It is shown\footnote{The statement $r\ge2$ there is an
    error: it originally
  referred to degree of differentiability, and has no connection with
  the restitution coefficient here.} 
in~\cite[Proposition 1]{DC} (see also~\cite{BD} where it is inherent
in the analysis, and see the calculations below) that
in a \nhd\ of $\Pi_0$ the impact surface
$V_c$ consists of two sheets (smooth 2-manifolds) intersecting
  transversely along $\Pi_0\,$.  One of these is of course the plane $\Pi$
while the other sheet $V_c'$ is represented in a \nhd\ of $\Pi_0$
as the graph of a smooth function $v=v_c(\tau,t)$  given implicitly by
\begin{equation}  \label{e:imp}
y_c(\tau,v_c;t)=0
\end{equation} 
when $x_c$ is written in the form
\begin{equation}  \label{e:yeq}
x_c=c+ty_c(\tau,v;t)
\end{equation}
for a smooth function $y_c$ on a \nhd\ of $\Pi_0$ in $\br^3$.
\medskip

By implicit differentiation of~(\ref{e:imp}) we find
\begin{equation}  \label{e:acc}
\dbd{}t\,v_c(\tau,t)|_{t=0}=-\tfrac12 a_c(\tau)
\end{equation}
where $a_c(\tau):=\ddot x_c(\tau,0;0)$ is the initial acceleration 
(again see~\cite{BD} as well as \cite[Proposition 1]{DC}).
\msk

There are three maps which we use to build the dynamical model 
for the impact oscillator.  We now describe each of these in turn.
%
%
\subsubsection{The projection $p_c$}
\begin{figure}[!ht]  
\begin{center}
 \scalebox{.60}{\includegraphics{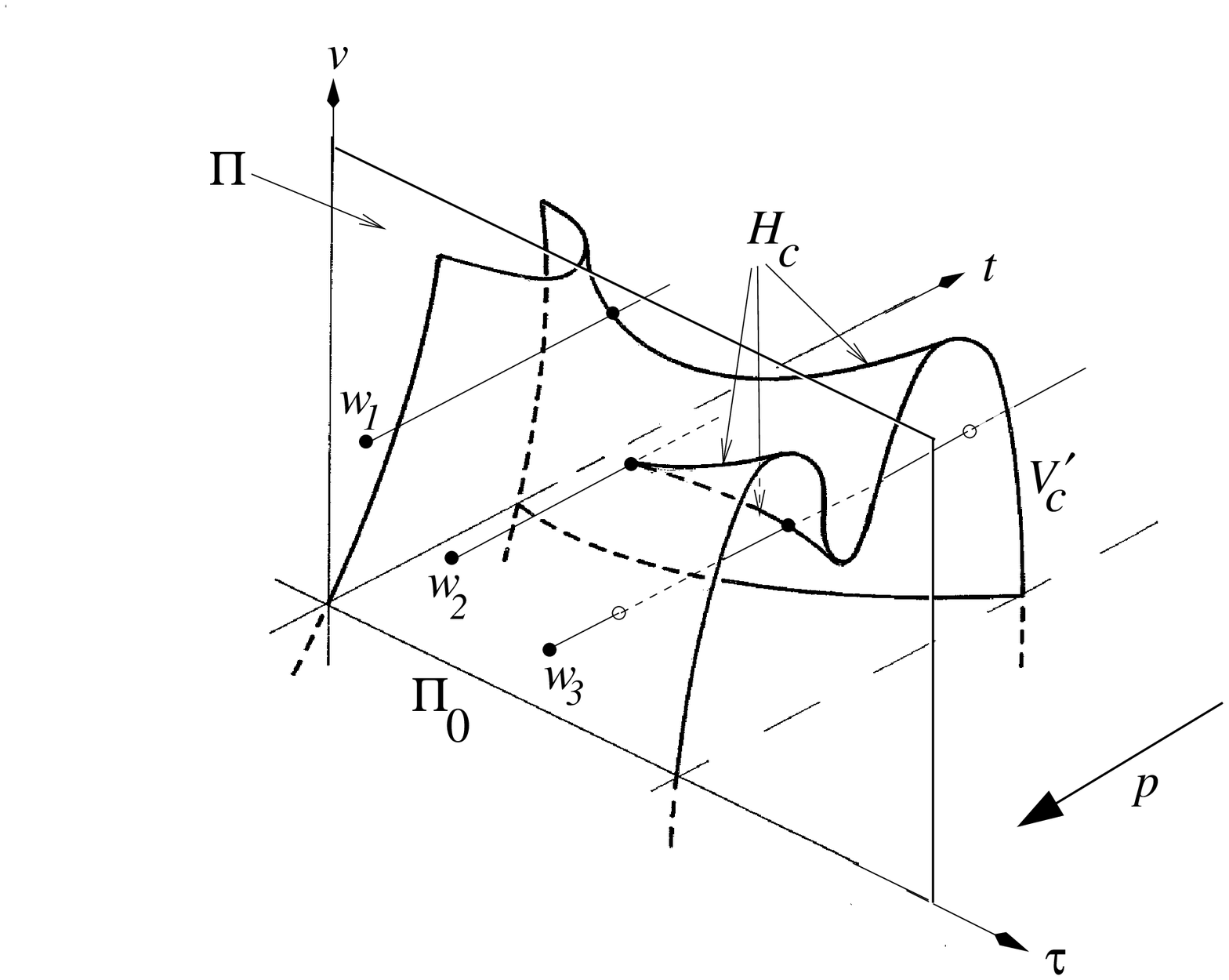}}
\end{center}  
\caption{The impact surface $V_c=\Pi\cup V_c'$ with horizon $H_c$
  that is the set of singular points of the projection
  $p|V_c'\to\Pi$.  Points $w_1,w_2,w_3$ belong to the apparent outline
  $P_c=p(H_c)$ (not shown).}  \label{fi:fig1}
\end{figure}

Consider the projection
\[
 p:\R^3\to\Pi : (\tau,v;t) \mapsto (\tau,v)
\]
from $\R^3$ to the plane $\Pi$ of initial data, and let $p_c$ denote
its restriction to $V_c'\,$:
\[
p_c:=p|V_c':V_c' \to \Pi.
\]
Let $H_c$ be the set of singular points of $p_c:V_c'\to\Pi$
  (including those, if any, where $V_c'$ itself may be singular). 
From the Implicit Function Theorem it is immediate that 
\[
H_c= \{(\tau,v;t)\in V_c' : \dot x_c(\tau,v;t)=0 \}
\]
which in geometric terms is the \emph{horizon} of $V_c'$ as viewed along
the $t$-direction.  Dynamically, the points of $H_c$ represent {\em graze}
points, that is points where the solution to~(\ref{e:eqn}) meets the
obstacle $x=c$ with zero velocity. See Figure~\ref{fi:fig1}, where for
diagrammatic simplicity the curve $H_c$ is not distinguished from the horizon of
$V_c'$ as seen from the viewpoint of the reader: in reality these
are of course different.
The subset $P_c:=p(H_c)$ of $\Pi$
is the \emph{apparent outline} (classically the \emph{apparent contour}) 
of the surface $V_c'$ in the $t$-direction: 
see Bruce~\cite{BR84b},Bruce {\em et al.}~\cite{BGx}.
\msk

The generic local geometry of smooth maps from a smooth surface $M$
into $\R^2$ is well understood since the work of Whitney~\cite{WH}.
The set of singular points is a smooth $1$-manifold in $M$ consisting
of arcs of {\em fold points} (where the singular set maps locally to a
smooth $1$-manifold) and isolated {\em cusp points} (where the
singular set maps locally to a $\frac32$-power cusp).  The precise
characterisations are given in terms of partial drivatives, and local
coordinates can be chosen so that at fold points or cusp points the
map has the form $(x,y)\mapsto(x,y^2)$ or $(x,y)\mapsto(x,xy+y^3)$
respectively.  These generic results apply also to the conceivably more
restricted setting of projection of a surface in
$\R^3$ to a plane, and thus to the geometry of apparent outlines.
In Figure~\ref{fi:fig1} the points $w_1,w_3$ are fold points, while
$w_2$ is a cusp point. 
\msk
  
\begin{figure}[!ht]   
\begin{center}
\scalebox{.50}{\includegraphics{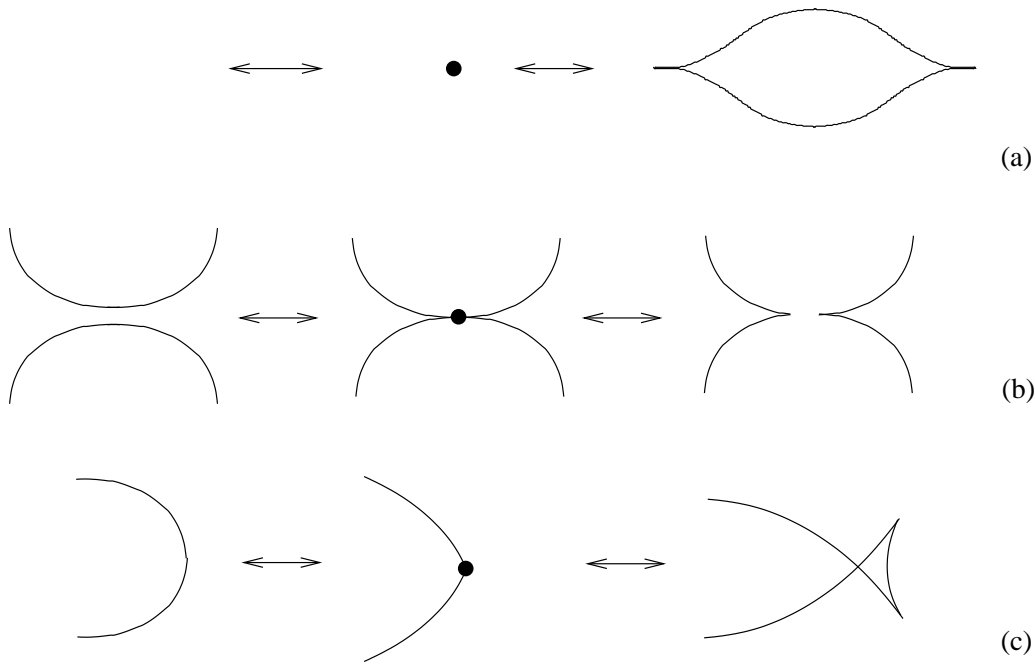}}
\end{center}
  \caption{The three generic transitions in apparent outlines:
    (a)~lips, (b)~beaks, (c)~swallowtail.}  \label{fi:fig2}
\end{figure}

As the surface or the view direction 
varies with one or more parameters the outline will typically undergo 
certain local transitions.  The generic 1-parameter transitions in
apparent outlines are also well understood: see
\cite{A},\cite{BR84a},\cite{BG85} and Figure~\ref{fi:fig2}.
\msk

In general, without any restriction on $t$,  there may be regions of
$\Pi$ filled densely by $P_c\,$.  However, if $V_c$ is periodic in $t$ 
(as can certainly happen for linear systems) or we restrict to a
compact subset of $V_c\,$ then this complication will not arise.
\msk

It is pertinent to ask whether for arbitrary choice of
clearance $\,c\,$ the apparent outline of the 
impact surface $V_c'\,$, which is determined by solutions of the
differential equation~(\ref{e:eqn}),  
is indeed generic as just described, and whether the
transitions in the apparent outline as $\,c\,$ varies are
also generic in this context.
These questions may be posed both for a general system~(\ref{e:eqn}) and for
a specific system such as the forced linear oscillator
\begin{equation}  \label{e:lin}
\ddot x + x = \cos\omega t.
\end{equation}
A detailed discussion of the generic geometry of $V_c'$ and its outline
for the general system~(\ref{e:eqn}) and for the linear system~(\ref{e:lin})
can be found in~\cite{DC} where these questions are answered.
\subsubsection{The re-set map $\phi_c\,$}
There is a natural smooth {\em re-set map} $\phi_c:V_c\to\Pi$ 
obtained simply by re-setting the clock to $t=0$ at the moment of impact:
\[
\phi_c:V_c\to\Pi : (\tau,v;t) \mapsto (\tau+t,\dot x_c(\tau,v;t)).
\]
Let $Z_c$ denote the {\em zero set} of $V_c$, that is the
intersection of $V_c'$ with the plane $v=0$.
The following facts about the singularity structure of $\phi_c$ are
proved in~\cite[Propositions 6,7]{DC}.
\begin{prop} \label{p:phising}
The singular set of $\phi_c$ is $Z_c\setminus\Pi_0$.  Points 
$(\tau,0;t)\in Z_c\setminus \Pi_0$
where $a_c(\tau)\ne0$ are fold points of $\phi_c$, while points 
where $a_c(\tau)=0$ but $a_c'(\tau)\ne0$ are cusp points.
\end{prop}
\subsubsection{The restitution map $R_c\,$}

The \emph{restitution map}
describes how the velocity immediately after impact depends on
the velocity immediately before impact.  Thus we write
\begin{eqnarray*}
\Pi^+=\{(\tau,v)\in\Pi:v>0\}\\
\Pi^-=\{(\tau,v)\in\Pi:v<0\}
\end{eqnarray*}
and we suppose that $R_c:\Pi^- \to \Pi^+$
is the restriction to $\Pi^-$ of a map   
\[
R_c:\Pi\to\Pi:(\tau,v)\mapsto(\tau,\rho_c(v)) 
\]
where $\rho_c:\br\to\br$ is a differentiable function with 
$\rho_c(0)=0$ and $\rho_c'(0)<0$.
In many cases an appropriate
choice is $\rho_c(v)=-rv$ with $0<r<1$ independent of $c$, 
although all that we shall require is that $\rho_c'(0)=-r$. 

\subsection{The first-hit map $F_c$ and the dynamical system $G_c$} \label{ss:hit}
Since we are assuming the dynamics take place in the region
$x\ge c$ it is only the upper half-plane $\Pi_0^+:=\Pi^+\cup\Pi_0$
that is of physical interest.  Nevertheless, in much of what follows
it will be useful first to consider $\Pi$ as a whole and then 
restrict to $\Pi_0^+$ later.
\msk

The projection $p_c:V_c\to\Pi$ has a partial right inverse
constructed as follows.
Choose initial data $(\tau,v)\in\Pi^+$ (so $x=c$ and $v>0$ at $t=0$), 
then proceed parallel to the $t$-axis in the direction of increasing $t$
until the point $F_c(\tau,v)$ of first intersection
(\emph{first hit}) with $V_c$ is reached: thus 
$$F_c(\tau,v)=(\tau,v;t_1)\in V_c'$$
where $(\tau,v;t)\notin V_c$  for $0<t<t_1$.
It is possible that no such $t_1$ exists (no future impacts occur),
but for notational simplicity we disregard such points and write
\[
F_c:\Pi^+\to V_c'
\]
in any case, with the understanding that $F_c$ may not be defined on
the whole of $\Pi^+$.
\msk

The construction of $F_c$ extends naturally to points $(\tau,0)\in\Pi_0$
for which $x_c(\tau,0;t)>0$ for all sufficiently small $t>0$; if this
does not hold (for example, if  $\ddot x_c(\tau,0;0)<0$)
then the constrained orbit remains stationary at $x=c$ 
and we define $F_c(\tau,0)=(\tau,0)$.
The points of $V_c'$ in the image of $F_c:\Pi_0^+\to V_c'$ will be
called {\em visible} points, as these are the points of $V_c'$ that can
be \lq seen\rq from $\Pi_0^+$ along the positive $t$ direction.

\medskip

Finally, although it is not neded for dynamical purposes,
for geometric convenience we define  $F_c:\Pi^-\to V_c'$ analogously,
proceeding here in the direction of negative $t$.
\msk

The first-hit map $F_c:\Pi\to V_c'$ defined in this way
typically has discontinuities. 
Since $p_c$ is a local diffeomorphism at points $z\in V_c'\setminus
H_c$ it follows that the discontinuities of $F_c$ must lie in 
the apparent outline $P_c=p(H_c)$.
Not all points $w\in P_c$ are discontinuity points of $F_c\,$, of course,
but only those with $F_c(w)\in H_c\,$: thus in Figure~\ref{fi:fig1}
the point $w_1$ is a point of discontinuity of $F_c$ while the point $w_3$
is not.
\subsection{The correspondence principle}
The task of describing the dynamics of the
impact oscillator close to grazing 
is greatly helped by exploiting a correspondence, induced naturally 
by $\phi_c\,$, between geometric features of
$P_c\,$ close to $\Pi_0$ and those away from $\Pi_0\,$.
\msk

If $z\in H_c$ then $p_c(z)\in P_c$ by definition, and since
$\dot x_c=0$ at $z$ we have $\phi_c(z)\in\Pi_0$. 
Likewise if $z\in Z_c$ then $p_c(z)\in \Pi_0$ by definition, and it is
also the case that $\phi_c(z)\in P_c\,$. This is because $z$
represents current data (on $x=c$) for a solution with initial
velocity $v=0$ so, by uniqueness of solutions of~({\ref{e:eqn}), after
re-set the point $\phi_c(z)$ represents initial data for a solution 
$x(t)$ with $x(t)=c$ and with $\dot x(t)=0$ at some other time $t\,$.   Thus 
\[
\phi_c(H_c)=\Pi_0=p_c(Z_c)
\quad\mbox{\ and\ }\quad
p_c(H_c)=P_c=\phi_c(Z_c). 
\]
In fact by extending this argument we obtain a more general result
which we leave as an exercise:
\begin{prop}  \label{p:corresp}
$$p_c^{-1}(P_c\cup\Pi_0)=\phi_c^{-1}(P_c\cup\Pi_0).$$
\endproof
\end{prop}
Furthermore, since re-setting the time clearly does not affect 
geometric features of the dynamics, any features of the outline $P_c$ 
in a \nhd\ of $w=p_c(z)$ where $z\in H_c$
must have counterparts in a \nhd\ of $\phi_c(z)\in\Pi_0$. 
This {\em correspondence principle} between $P_c$ and $\Pi_0$
enables us to focus the analysis on points of
$\Pi_0$ when studying local dynamics close to grazing, and (as we see
below) provides a valuable framework for understanding the structure of the
dynamics itself. A fuller description of
the geometry associated to the correspondence principle can be found in~\cite{DC}.
\subsection{Construction of the dynamics}
We are now ready to construct the (discontinuous) discrete dynamical
system on the half-plane $\Pi_0^+$ that we will use to model the dynamics 
of the impact oscillator.
\medskip

Begin by applying the first-hit map $F_c:\ppb\to V_c'$. This corresponds
to following a solution of~(\ref{e:eqn}) as it leaves the obstacle at
$x=c$ until it returns to $x=c$ for the first time.  Next, apply the
re-set map $\phi_c:V_c'\to\Pi$; this is
simply a re-calibration of the time variable.
Since the initial data lie in $\ppb$ the velocity at first hit must be
negative or zero, so we have 
\[
\phi_c\circ F_c:\ppb\to\pnb:=\Pi^-\cup\Pi_0.
\]
Finally apply the restitution map $R_c:\pnb\to\ppb\,$:
\[
\ppb \overset{F_c}\rightarrow V_c' \overset{\phi_c}\rightarrow\pnb 
                                     \overset{R_c}\rightarrow\ppb\,.
\]

\medskip

The iteration of this 
process reconstructs the impact dynamics of the oscillator as a
(discontinuous) discrete dynamical system on $\ppb$ generated by the map
\[
G_c:=R_c\circ\phi_c\circ F_c : \ppb \to \ppb.
\]
As indicated above, we may regard $F_c$ and $R_c$ as maps defined on
the whole of $\Pi$, and hence the same applies to $G_c\,$ whenever this
is convenient.
\msk

The rest of this paper is devoted to describing fully the generic 
behaviour of iterations of $G_c$ close to the $\tau$-axis $\Pi_0\,$, 
both for fixed clearance $c$ and as a 1-parameter family of dynamical
systems when $c$ or another parameter is varied. 
We take the partial results of~\cite{BD} further by giving a rigorous
derivation of the dynamics for nondegenerate chatter at 
{\em regular points} of $\Pi_0$ (that is where $a_c(\tau)\ne0$)
and degenerate chatter at {\em tangency points} of $\Pi_0\,$
(where $a_c(\tau)=0$ but the derivative $a_c'(\tau)\ne0$). 
We also consider a yet more degenerate situation that generically arises
only for certain values of the clearance $c$ at so-called {\em swan points} 
of $\Pi_0\,$ (where $a_c(\tau)=a_c'(\tau)=0$ but $a_c''(\tau)\ne0$), 
and we show how the dynamics unfold and undergo
newly-created discontinuities in a $1$-parameter bifurcation  
as $c$ passes through such values. 
The creation of new cusp points for $p_c$ gives rise to
convoluted stable manifolds of nearby chatter points, with immediate
implications for complicated global dynamical behaviour.
Understanding the global structure of stable manifolds is one of the
main challenges in formulating a dynamical theory for non-smooth
systems such as impact oscillators.

\msk
\section{Nondegenerate chatter} \label{s:nondegen}
Let the clearance $c$ be fixed. To simplify notation in this section
we temporarily write $V',F,G$ etc. for $V_c',F_c,G_c$ etc., although we
retain $v_c$ for clarity and understand that $p$ denotes $p|V_c'$.
Using coordinates $(\tau,v)$
on $\Pi$ and $(\tau,t)$ on $V'\,$,
we identify a point $(\tau_*,0;0)$ of the $\tau$--axis $\Pi_0=\Pi\cap V'$
with both $w_*=(\tau_*,0)\in\Pi$ and $z_*=(\tau_*,0)\in V'\,$. 
\msk

Let $w_*=(\tau_*,0)\in\Pi_0$ be a {\em regular point} of $\Pi_0\,$,  
so that $a(\tau_*)\ne0$. 
Since the \lq free\rq\ dynamics takes place in the region
$x>c\,$, the behaviour of any solution that leaves $x=c$ with zero velocity
but positive acceleration will not be susceptible to our local
(low-velocity) analysis close to $\Pi_0\,$: therefore we assume 
$a_*:=a(\tau_*)<0\,$. On leaving $x=c$
with small positive velocity the solution to~(\ref{e:eqn}) will slow
down, then possibly reverse and hit $x=c$ again after a short time.
This is indeed the case as we see from~(\ref{e:acc}): since 
$z_*\notin H$ the map $p:V'\to\Pi$ is a local diffeomorphism at $z_*$ and
therefore $F$ is also a local diffeomorphism at $w_*\,$.
From Proposition~\ref{p:phising} we have that $\phi:V'\to\Pi$ is a
local diffeomorphism at $w_*$ and so $G=R\circ\phi\circ F$
is (the restriction to $\Pi_0^+$ of) a local diffeomorphism of $\Pi$ at $w_*\,$.
\subsection{Local linearisation along $\Pi_0$}
As a first approximation to the dynamics we linearise at the
equilibrium point $w_*\in\Pi_0$.  
Since by definition
$$G\circ p=R\,\circ\phi$$
we have on differentiation
$$DG(w)Dp(z)=DR(u)D\phi(z)$$ 
for $z\in V'$ where $w=p(z)$ and $u=\phi(z)$, both in $\Pi$.
Writing $\tau=\tau_*+\sig$ we see from~(\ref{e:acc}) 
that the derivative $Dp(z_*)$ is  
\[
Dp(z_*):(\sig,t)\mapsto(\sig,-\tfrac12 a_*t).
\]
Also, using implicit differentiation of~(\ref{e:imp}) to see that 
$\dbd{v_c}t(z_*)=\tfrac12a_*$
we find (see~\cite[Lemma 2]{DC})
\[
D\phi(z_*):(\sigma,t)\mapsto(\sigma+t,\tfrac12a_* t).
\]
Hence the derivative of the local diffeomorphism 
$G=R\circ\phi\circ F$ at $w_*$ is given by the $2\times 2$ matrix
\begin{equation}
DG(w_*) = 
\left( 
\begin{array}{ccc}  
1 & -\tfrac2{a_*}  \\  
0 & r
\end{array}
\right) 
\end{equation}
where $r=-\rho'(0)$ is the restitution coefficient
(compare~\cite{BD}). 
\msk

Eigenvectors of $DG(w_*)$ are $(1,0)$ with
eigenvalue $1$ (since the $\tau$-axis $\Pi_0$ is fixed) 
and $(2,a_*(1-r))$ with eigenvalue $r$.
Therefore, to first order, we have the picture as illustrated in
Figure~\ref{fi:fig3}(a) (see also~\cite{BD}): 
orbits of the local linearisation of the
discrete dynamics $G$ approach $\Pi_0$ along directions with negative
slope, and this slope tends to zero as $a_*\to0$ or as $r\to1$.
\begin{figure}[!ht]  
  
\begin{center}
\scalebox{0.55}{\includegraphics{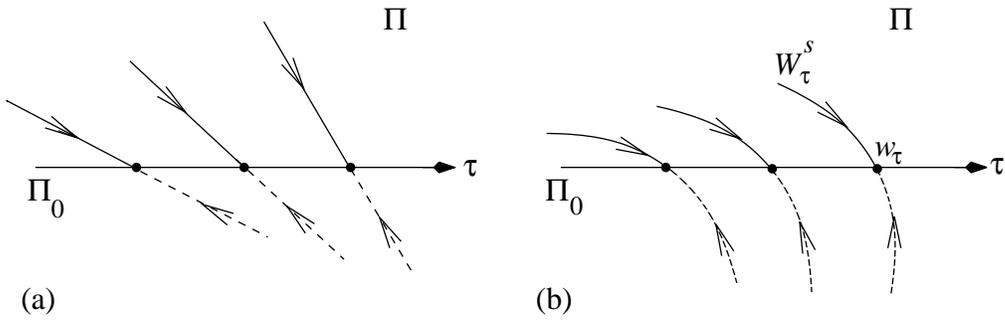}}
\end{center}
  \caption{Local dynamics of $G$ in the $(\tau,v)$-plane $\Pi$
   close to $\Pi_0\,$: nondegenerate chatter for $a(\tau)<0$. 
   (a) Local linearisation, (b) local foliation by stable manifolds.}
  \label{fi:fig3}
\end{figure}
\msk

This does not complete the story, however, as it remains to confirm that
this linear description also accurately describes the nonlinear
behaviour of $G$ on $\Pi$ close to $\Pi_0$. 
\medskip

A fundamental component in the theory of smooth dynamical systems 
is the analysis of dynamics in the \nhd\ of a
 {\em normally  hyperbolic} invariant manifold, 
that is an invariant manifold for which the rate of normal
contraction or expansion overpowers any contraction or expansion
along the manifold itself: main references are~\cite{HPS} and~\cite{TK}. See
also~\cite{HP} for an encyclopaedic treatment.  

In our case, let $J$ be any open interval of the $\tau$-axis $\Pi_0$
on which $a(\tau)$ is negative and bounded away from zero.  Then $G$
is the identity on $J$ and at each point $w\in J$ the
derivative $DG(w)$ has a contracting eigenspace (since $0<r<1$) transverse to
$\Pi_0$.  Hence $J$ is normally hyperbolic for $G$, and 
from the standard theory~\cite{HPS} we immediately deduce the 
following description of the local dynamics of $G$ on some \nhd\ $U$
of $J$ in $\Pi$.  Recall, however, that for appplications to the
impact oscillator we are concerned only with $\ppb$, that is where $v\ge0$.
\begin{theo}  \label{t:t1}
For each $\tau\in J$ there is a smooth $G$-invariant curve $W^s_\tau$ 
in $U\subset\Pi$ through $w_\tau=(\tau,0)\in\Pi_0$  
and tangent to the direction $(2,(1-r)a(\tau))$.
The curve $W^s_\tau$ is a local {\em stable manifold} for $w_\tau$:
\[
W^s_\tau=\{u\in U:G^n(u)\to w_\tau \mbox{\ as\ } n \to +\infty\}.
\]
The curves $\bigl\{W^s_\tau\bigr\}_{\tau\in J}$ provide a continuous
foliation of $U$, in that there is a homeomorphism
$\Phi:J\times\br\to U$ taking $\{\tau\}\times\br$ to $W^s_\tau$ for each
$\tau\in J$. See Figure~\ref{fi:fig3}(b).
\end{theo}
Thus the linear picture is indeed an accurate representation of the
nonlinear dynamics close to $\Pi_0\,$, although this result
yields only that the stable manifold $W^s_\tau$ varies {\em continuously}
with $\tau$. Can we obtain more, and see that in fact $W^s_\tau$ varies
{\em smoothly} with $\tau$ so that the union of the $W^s_\tau$ form a {\em smooth
foliation} of a \nhd\ of $\Pi_0\,$? From a result of
Takens~\cite{TK} the answer turns out to be positive.
\begin{theo}
If the functions $f$ and $g$ in (\ref{e:eqn}) are $C^\infty$ then (for
suitable $U$) the homeomorphism $\Phi$ in Theorem~\ref{t:t1} may be
taken to be a $C^\infty$ diffeomorphism. 
\end{theo}
\proof
This follows from the more general (although local) theorem
in~\cite{TK} concerning partially hyperbolic fixed points.  
The {\em non-resonance} conditions which are imposed
in the more general setting of~\cite{TK} are in our case vacuously
satisfied since $0<r<1$ and there is no expansion or contraction along
the manifold $J$.  Extension from the local result to the whole of $J$
follows by usual compactness arguments applied to the closure of $J$.
\endproof 
\msk

Therefore the dynamics of the local linearisation as described in~\cite{BD}
do indeed give an accurate picture of the true (nonlinear) dynamics up
to smooth change of coordinates.
\bigskip

As $a(\tau)\to0$ the stable manifold $W_\tau^s$ approaches $\Pi_0$ with 
slope tending to zero.  Our next step is to investigate the local dynamics
in this limiting case.
\section{Degenerate chatter}  \label{s:degen}
In this section we examine the nature of chatter close to a point
$w_*=(\tau_*,0)$ where now $a_c(\tau_*)=0$.  For generic choice of the
functions $f$ and $g$ in~(\ref{e:eqn}) and for an open dense set of
choices of the clearance $c$ the $T$-periodic function $a_c(\tau)$ will
have (modulo T) a finite set of zeros each with $a_c'(\tau)\ne0$.
In particular this is the case for the linear system~(\ref{e:lin}),
since here $a_c(\tau)=-c+\cos\omega\tau$ where $T=2\pi/\omega$.  However,
for certain values of $c$ the function $a_c(\tau)$ can typically be
expected to undergo Morse transitions (fold bifurcations) 
where $a_c(\tau)=0$ and $a_c'(\tau)=0$ while $a_c''(\tau)\ne0$.
We shall describe the local dynamics of the impact oscillator in both
these situations.
\medskip

We start by calculating the first few terms of the Taylor series of
the maps
\begin{equation} \label{e:proj}
p_c:(\tau,t)\mapsto (\tau,v_c(\tau,t))
\end{equation}
and
\begin{equation} \label{e:phi}
\phi_c:(\tau,t)\mapsto (\tau+t,\dot x_c(\tau,v_c(\tau,t))
\end{equation}
at points $(\tau,0)\in\Pi_0\,$.  
Since from (\ref{e:yeq}) we have
\begin{equation}  \label{e3}
v=\dot x_c(\tau,v;0)=y_c(\tau,v;0)
\end{equation}
it follows that  $\dbd{y_c}v=1$ and all other derivatives of $y_c$ with
respect to $v$ and/or $\tau$ vanish on $\Pi_0$.  Using this, by
implicit differentiation of~(\ref{e:imp}) and writing $\tau=\tau_*+\sig$
we find
\begin{eqnarray}  \label{e:vtaylor}
v_c(\tau_*+\sig,t) = &-\tfrac12 a t  \\
          &\quad-\tfrac12 a'\sig t+\tfrac12(-\tfrac13 a'+\tfrac12 a b)t^2  \\
     &\qquad-\tfrac14 a''\sig^2t +\tfrac12(-\tfrac13a''+\tfrac12a'b)\sig t^2 \\
&\qquad\quad-\tfrac16(\tfrac14a''+\tfrac32(-\tfrac13a'+\tfrac12ab)b)t^3 +O(4)
\end{eqnarray}
where $a=a_c(\tau_*)$,
$\,b=\dbd{}v\ddot x_c(\tau_*,0;0)$
and the primes denote derivatives with respect to $\tau$ evaluated at
$\tau_*\,$ (which in general depend also on $c\,$).  
Hence, if $a\ne0$ then
\begin{equation}
v_c(\tau_*+\sig,t)= -\tfrac12 a t  + O(2)
\end{equation} 
while if $a=0$ but $a'\ne0$ then
\begin{equation} \label{e:degen}
v_c(\tau_*+\sig,t)= -\tfrac16a't(3\sig + t) + O(3).
\end{equation}
Finally, if $a=a'=0$ but $a''\ne0$ then
\begin{equation} \label{e:swan}
v_c(\tau_*+\sig,t)= -\tfrac1{24} a''t(6\sig^2+4\sig t + t^2) + O(4).
\end{equation}  
\subsection{Dynamics near a tangency point} \label{ss:tang}
A point $(\tau,0)\in\Pi_0$ where 
$a_c(\tau)=0$ and $a_c'(\tau)\ne0$ is called a {\em tangency point}.  
In the context of Section~\ref{s:nondegen}
this can be regarded as a point of $\Pi_0$ where
the normal hyperbolicity fails in the least degenerate way.

Suppose that $w_*=(\tau_*,0)\in\Pi_0$ is a tangency point.
From~(\ref{e:degen}) we see that $v_c$ has a saddle critical point at 
$(\sig,t)=(0,0)$.  The horizon $H_c$ is tangent to the line $3\sig+2t=0$
which projects from $V_c'$ by $p_c$ to the parabola
$v=\tfrac38 a_c'(\tau_*)\sig^2$ in $\Pi$.  We consider two cases separately,
according to the sign of $a_c'(\tau_*)$.
Again for convenience we drop the suffix $c$ from the notation.
\subsection*{(a) The case $a'(\tau_*)>0$}
The points of $H$ close to $w_*$ have $v\ge0$.  The part of $H$ relevant
for us is the arc where $t\ge0$, that is $\sig\le0$, and so the
relevant part of $P=p(H)$ here is a smooth arc $P_*$ 
quadratically tangent to $\Pi_0$ at $w_*$ from the direction
$\tau<\tau_*$ and $v>0$. This is the discontinuity set for the
first-hit map $F$ and hence for $G$ locally: we call it the {\em
discontinuity arc}.  See Figure~\ref{fi:fig5}.

\begin{theo}   \label{t:t2}
Let $w_*=(\tau_*,0)$ be a tangency point with $a'(\tau_*)>0$.   
The map $G$ has a unique invariant arc $\gamma\subset\Pi_0^+$ with
an end-point at $w_*$, this arc being smooth and quadratically tangent
to $\Pi_0$ at $w_*$ from the side $\tau<\tau_*$ and lying between
$P_*$ and $\Pi_0$.
The fixed point $w_*$ is exponentially attracting for $G|\gamma$, while
$\gamma$ is weakly normally hyperbolic (repelling) for $G$.
Orbits close to but not on $\gamma$ either (if below $\gamma$)
are attracted exponentially 
to $\Pi_0$ in nondegenerate chatter or (if above $\gamma$) lose
contact with the obstacle on crossing the discontinuity curve $P_*$
after finitely many impacts.  See Figure~\ref{fi:fig5}.
\end{theo}
By \emph{weakly normally hyperbolic} we mean that at points
$w\in\gamma$ the derivative $DG(w)$ has an eigenvalue $\lambda\ne1$
(in this case $\lambda>1$) with eigendirection transverse to
$\gamma$.  However, $\lambda\to1$ as $w\to w_*\,$ along $\gamma$.
The terms {\em above} and {\em below} correspond to values of $v$.
\msk

\begin{figure}[!ht]  
\begin{center} 
 \scalebox{0.5}{\includegraphics{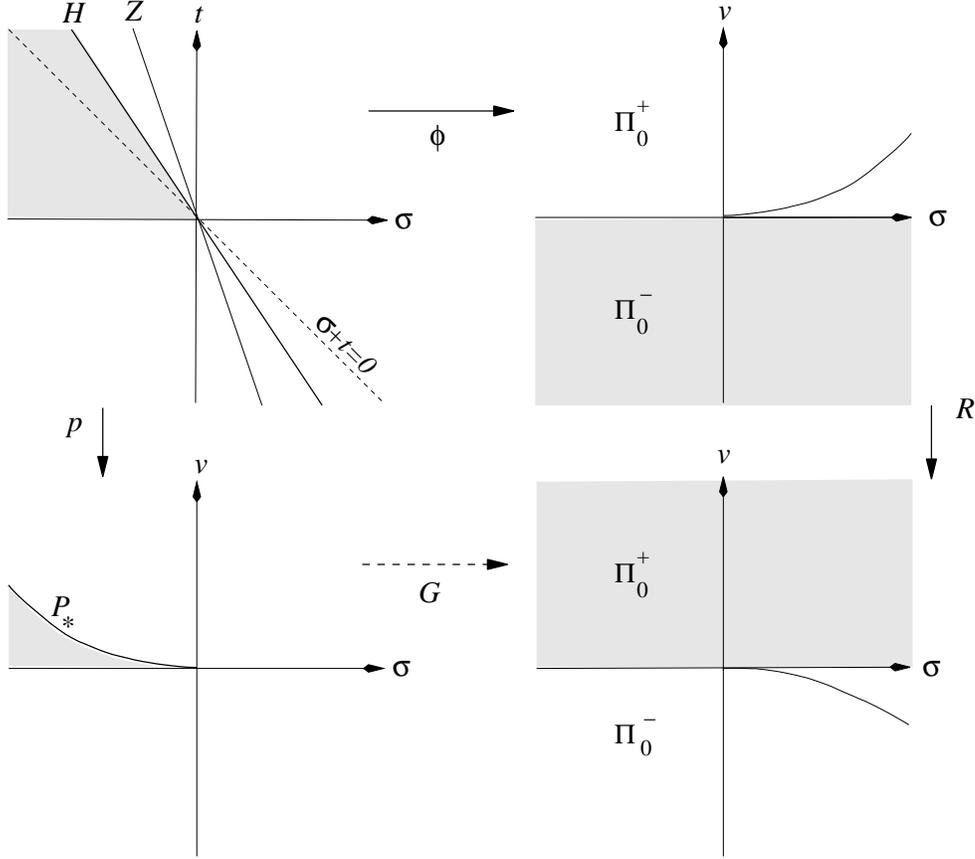}}
\end{center}  
\caption{The actions of $p$,$\,\phi$ and $R$ close to $(\tau_*,0)$ 
           with $a'(\tau_*)>0$: here $\tau=\tau_*+\sig$. The shaded
           area in the top left diagram represents the visible part of
           $V'$, and the other shaded areas are its images under the
           relevant maps as indicated.} 
   \label{fi:fig4}
\end{figure}
A proof of the existence of such an invariant curve $\gamma$ was
outlined by Budd and Dux in~\cite{BD}
by restricting to second-order terms. 
The analysis in~\cite{BD} is also for a particular linear system, although
the authors observe that their results apply more generally.
We provide here a rigorous
proof that applies in full generality to nonlinear systems of type~(\ref{e:eqn}).
\subsubsection{Proof of Theorem~\ref{t:t2}}
From (\ref{e:proj}) and (\ref{e:degen}) we see that the
projection $p_c:V'\to\Pi$ is given in local coordinates by
\begin{equation}  \label{e:pnew}
p:(\sig,t) \mapsto (\sig, -k(3\sig t+t^2) + O(3))
\end{equation}
where $k=\tfrac16a'(\tau_*)>0$, and from~(\ref{e:phi}) and~(\ref{e:degen})
the reset map $\phi:V'\to\Pi$ is given by
\begin{equation}  \label{e:phinew}
\phi:(\sig,t)\mapsto (\sig+t,-k(3\sig t+2t^2) + O(3)).
\end{equation}
In Figure~\ref{fi:fig4} we indicate some local geometry of $p$ and
of $\phi$.  As already noted, the horizon $H$ is given by $3\sig+2t=0$, 
while to first order the zero set $Z$ is given by $3\sig+t=0$. 
Also, the line $\sig+t=0$ is taken by $\phi$ to the $v$-axis $\sig=0$.
\msk 

\begin{figure}[!ht]  
\begin{center}
  \scalebox{0.8}{\includegraphics{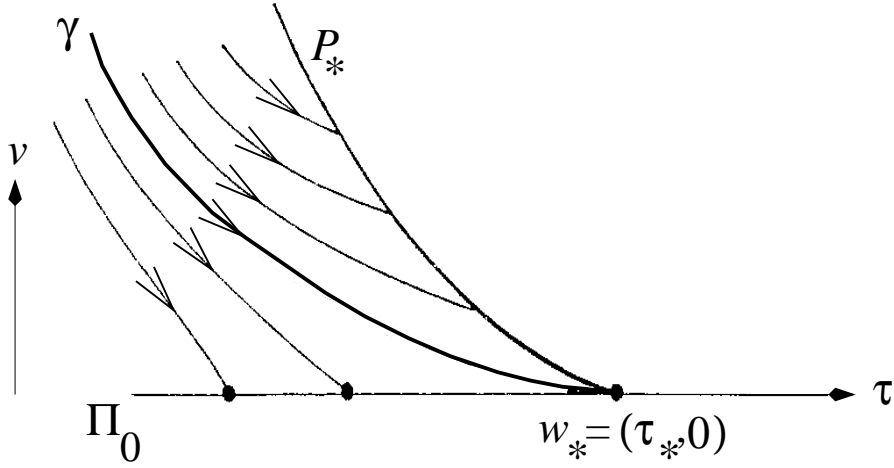}}
\end{center}  
\caption{Degenerate chatter at $w_*=(\tau_*,0)$ with $a'(\tau_*)>0$.  
  Invariant curves
  below the invariant curve $\gamma$ represent nondegenerate chatter;
  invariant curves above $\gamma$ carry orbits across the
  discontinuity arc $P_*$.}  \label{fi:fig5}
\end{figure}
Now take new coordinates $(\sig,\eta)$ in $\R^2$, where $\eta=\frac32\sig+t$. 
Then 
\[
p:(\sig,\eta) \mapsto (\sig,-k \eta^2 + \tfrac94 k \sig^2 +O(3)).
\]
Composing with the diffeomorphism
$$\psi:\R^2\to\R^2:(u,w)\mapsto(u,w-\tfrac94 ku^2)$$
we obtain
\[
\psi\circ p:(\sig,\eta)\mapsto(\sig,-k\eta^2 + O(3))
\]
which exhibits $p$ explicitly as a {\em fold} map.  By standard
methods of singularity theory (see e.g. Arnol'd {\em et al.}~\cite{AGV},
Golubitsky {\em et al.}~\cite{GG}), after composing with
$C^\infty$ diffeomorphisms of $\R^2$
in range and domain that are the identity to second order, we may in
fact assume that
\begin{equation} \label{e:pdef}
\psi\circ p:(\sig,\eta)\mapsto(\sig,-k\eta^2)
\end{equation}
without any higher order terms. 
\msk

Let $q:V'\to\Pi$ denote the composition $R\circ\phi$.
In the new $(\sig,\eta)$ coordinates we find 
\begin{equation}  \label{e:defq}
\psi\circ q:(\sig,\eta)\mapsto (\eta-\tfrac12\sig,
    -rk(2\eta^2-3\sig \eta)-\tfrac94k(\eta-\tfrac12\sig)^2 + O(3))
\end{equation}
where we recall that $r=-\rho'(0)$.
\medskip

To say that $\gamma$ is an invariant curve for $G$ is equivalent to
saying that the curve $\beta=F(\gamma) \subset V'$ has the property
that $p(\beta)=q(\beta)$, which in turn is equivalent to saying
that $\beta$ is an invariant curve for
the map $h:=p^{-1}\circ q$ whose domain and range are subsets of $V'$. 
Explicitly we have
\begin{equation} \label{e:hdef}
h=p^{-1}\circ q: (\sig,\eta) \mapsto
\bigl(\,\eta-\tfrac12\sig,\sqrt{Q_r(\sig,\eta)+O(3)}\,\bigr)
\end{equation}
where 
\begin{equation}  \label{e:qdef}
Q_r(\sig,\eta):= r(2\eta^2-3\sig \eta)+\tfrac94(\eta-\tfrac12\sig)^2.
\end{equation}
Our strategy for finding $\beta$ will be to blow up the coordinates
at $(\sig,\eta)=(0,0)$ and then exhibit
$\beta$ in the new coordinates as the stable manifold of a hyperbolic
fixed point of $G$.
\medskip

For the blow-up construction we write
\[
(\sig,\eta) = (s,sm) =: \chi(s,m)
\]
so that $m$ is the slope of the ray from the origin on which $(\sig,\eta)$
lies, and $s$ measures displacement along the ray.  Then
\[
h(s,sm)=\bigl(\,(m-\tfrac12)s,\sqrt{s^2Q_r(1,m)+O(s^3)}\,\bigr) 
\]
and so using $(s,m)$ as coordinates this becomes
\[
h':(s,m) \mapsto (s',m')
\]
where
\begin{eqnarray}
s'& =(m-\tfrac12)s \\
m'& =(m-\tfrac12)^{-1}\sqrt{Q_r(1,m)+O(s)}.
\end{eqnarray}   
The domain $D_r$ of $h$ is the subset of $\br^2$ corresponding to
the values of $m$ for which 
$m\ne\tfrac12$ and the expression under the square root is nonnegative.
The following is straightforward to check.
\begin{prop}
The quadratic function $m\mapsto Q_r(1,m)$ has zeros at $m_-,m_+$ where
\[
0 < m_-	< \tfrac12 < m_+ < \tfrac32  
\]
and $Q_r(1,m)>0$ for  $m\in E_r:=(-\infty,m_-]\cup [m_+,\infty)$.
Also  $m_-,m_+ \to \frac12$ as $r\to0$.
\endproof
\end{prop}
Since distinct zeros of a quadratic function are necessarily simple
zeros, the Implicit Function Theorem yields the following result.
\begin{cor}
For sufficiently small $\epsilon>0$ the domain $D_r$ of $h'$ consists
of the region $D_r(\epsilon)=D_r^-(\epsilon)\cup D_r^+(\epsilon)$ where
\begin{eqnarray}
D_r^-(\epsilon) &:= \{(s,m)\in\br^2 : |s|<\epsilon, m<\delta_-(s)\}, \\
D_r^+(\epsilon) &:= \{(s,m)\in\br^2 : |s|<\epsilon, m>\delta_+(s)\}
\end{eqnarray}
and $\delta_\pm$ are smooth functions with $\delta_\pm(0)=m_\pm$
respectively.
\end{cor}
 We now seek a fixed point for $h'$ on $0\times E_r$.  This occurs where
\[
m=(m-\tfrac12)^{-1}\sqrt{Q_r(1,m)},
\]
that is where
\begin{equation}  \label{e:fixm}
Q_r(1,m)=m^2(m-\tfrac12)^2.
\end{equation}
Solutions of (\ref{e:fixm}) are given by
\(
m=\tfrac32
\)
and by the solutions of
\begin{equation} \label{e:meq}
2rm=(m-\tfrac12)^2(m+\tfrac32).
\end{equation}
A simple sketch of the graph of the cubic function on the right hand
side of (\ref{e:meq}) shows that this equation has three solutions
$m_1,m_2,m_3$ with 
\[
m_1 < -\tfrac32 \quad\mbox{\ and\ }\quad 0 < m_2 < \tfrac12 < m_3.
\]
We now investigate these four solutions of (\ref{e:fixm}) and verify
that only one of them is relevant here.
\begin{description}
\item[Solution $m=\frac32$]: in the original $(\sig,t)$ coordinates this
corresponds to $t=0$, that is the $\tau$-axis $\Pi_0$.  Clearly
this is invariant under $G$ (on one side of $(\tau_*,0)$) but it represents
equilibrium states and is not the arc $\gamma$ representing chatter 
that we seek. In fact $m=\frac32$ implies $r=1$ which is the ideal case
of perfect restitution in the limit as $v\to 0$. 
\item[Solution $m_1<-\frac32$]: the image of this line is given
opposite orientations by $p$ and $q$, so that under the
composition $q\circ p^{-1}$ the orientation of the image is reversed.
Such a curve cannot be a candidiate for $\gamma$.  
Moreover, since the gradient of the line is less than $-3$ 
it maps under $p$ (or $q$) to a curve tangent to the $\tau$-axis but 
with $v\le0\,$, not relevant for us. 
\item[Solution $0<m_2<\frac12$]: 
the image of this line is also given
opposite orientations by $p$ and $q$ 
(in particular, points with $v>0$ are taken to points with $v<0$)
so this image cannot be a candidate for $\gamma$.
\item[Solution $m_3>\frac12$]: here when $t>0$ both
$p$ and $q$ take this line into the region $v>0$
with the same orientation, so if $\gamma=p(\beta)$ exists as an
invariant curve then the tangent direction to $\beta$ at the origin
must correspond to this ray.  

Note that $m_3$ increases from $\frac12$ to
$\frac32$ as $r$ increases from $0$ to $1$.
\end{description}
We next study local linearisation of $h'$ at $(0,m_3)$.
Using~(\ref{e:hdef}) and~(\ref{e:qdef}) we find that
the Jacobian matrix for $h'$ at a point $(0,m)$ has the form
\[
Dh'(0,m) =  \begin{array}{cc}
		m-\frac12 & 0 \\
		*         & N_r(m)
	   \end{array}
\]
with
\begin{eqnarray}
  N_r(m) &= Q^{-\tfrac12}
\bigl(\tfrac12(m-\tfrac12)^{-1}[r(4m-3)+\tfrac92(m-\tfrac12)] -
Q(m-\tfrac12)^{-2}\bigr)   \\
         &= r(2m+3)(2m-1)^{-2} Q^{-\frac12}
\end{eqnarray}
where $Q$ stands for $Q_r(1,m)$ and where $*$ denotes a term which depends
on the $O(3)$ term in $\phi$.   In the case when $m$ satisfies the
fixed point equation (\ref{e:fixm}) this further simplifies (eliminating
$r$) to give
\[
  N_r(m) = \frac{(2m+3)^2}{8m^2(2m-1)}.
\]
As the eigenvalues of $Dh'(0,m)$ are $\lam=(m-\tfrac12)$ and $\lam'=N_r(m)$
we easily conclude the following.
\begin{prop}
The fixed point $m_3$ with $\frac12<m_3<\frac32$ has eigenvalues
$\lam,\lam'$ with $0<\lambda<1$ and $\lam'>1$.  Also
\[
\lambda\to1,\quad \mu\to1\mbox{\qquad as\qquad }r\to1 \quad (m_3\to\tfrac32)
\]
while 
\[
\lambda\to0, \quad \mu\to\infty\mbox{\qquad as\qquad}r\to0\quad
  (m_3\to\tfrac12).
\]
The eigendirection for $\mu$ is the $m$-axis. 
\endproof
\end{prop}
Note that the eigendirection for
$\lambda$ would be the $s$-axis if there were no $O(3)$ terms in
$\phi$. 
\msk

Since $(0,m_3)$ is hyperbolic it has a unique smooth stable manifold
$W^s$ transverse to the $m$-axis.  
Then $\chi(W^s)$ is a smooth curve $\beta$ through the origin in $\br^2$ and
tangent to the $s$-axis from above ($\eta>0$) or below ($\eta<0$) according
as $\sig<0$ or $\sig>0$, and the same holds in the original $(\sig,t)$
coordinates. By construction, the invariance of $W^s$ under $h'$
implies invariance of $\beta$ under $h$, which in turn implies
invariance of
$\gamma:=p(\beta)\cap\ppb$ under $G$.  The dynamical
properties of $h'$ near the saddle point $(0,m_3)$ carry over via $\chi$ to the
stated properties of $G$ on and close to $\gamma$.
This completes the proof of Theorem~\ref{t:t2}.
\endproof
\subsection*{(b)\quad The case $a'(\tau_*)<0$}
Here the horizon $H$ for $(\tau,t)\ne(\tau_*,0)=w_*$ 
locally lies in the region $v<0$ and so 
$F:\ppb\to V'$ and hence also
$G:\ppb\to\ppb$ are continuous maps in a \nhd\ of $w_*$ in $\ppb$.  In
detail, we see that to first order $F$ maps a \nhd\ $\T$ of $w_*$ in
$\ppb$ onto a \nhd\  
of the origin in the sector 
$$\T:\quad 3\sig+t\ge0\,,\,t\ge0$$
in $\R^2$ (coordinates on $V'$\,) which is then taken by $\phi$ to the region
\[
\T'':\quad \tfrac38 a'(\tau_*)\sig^2\le v\le 0\ ,\quad \sig\ge0
\]
in $\ppb$: see Figure~\ref{fi:fig6}.  Finally, $R$ maps $\T''$ to a
similar if narrower region $R\T''$ in $\Pi_0^+$.
\begin{figure}[!ht]  
\begin{center}
\scalebox{0.5}{\includegraphics{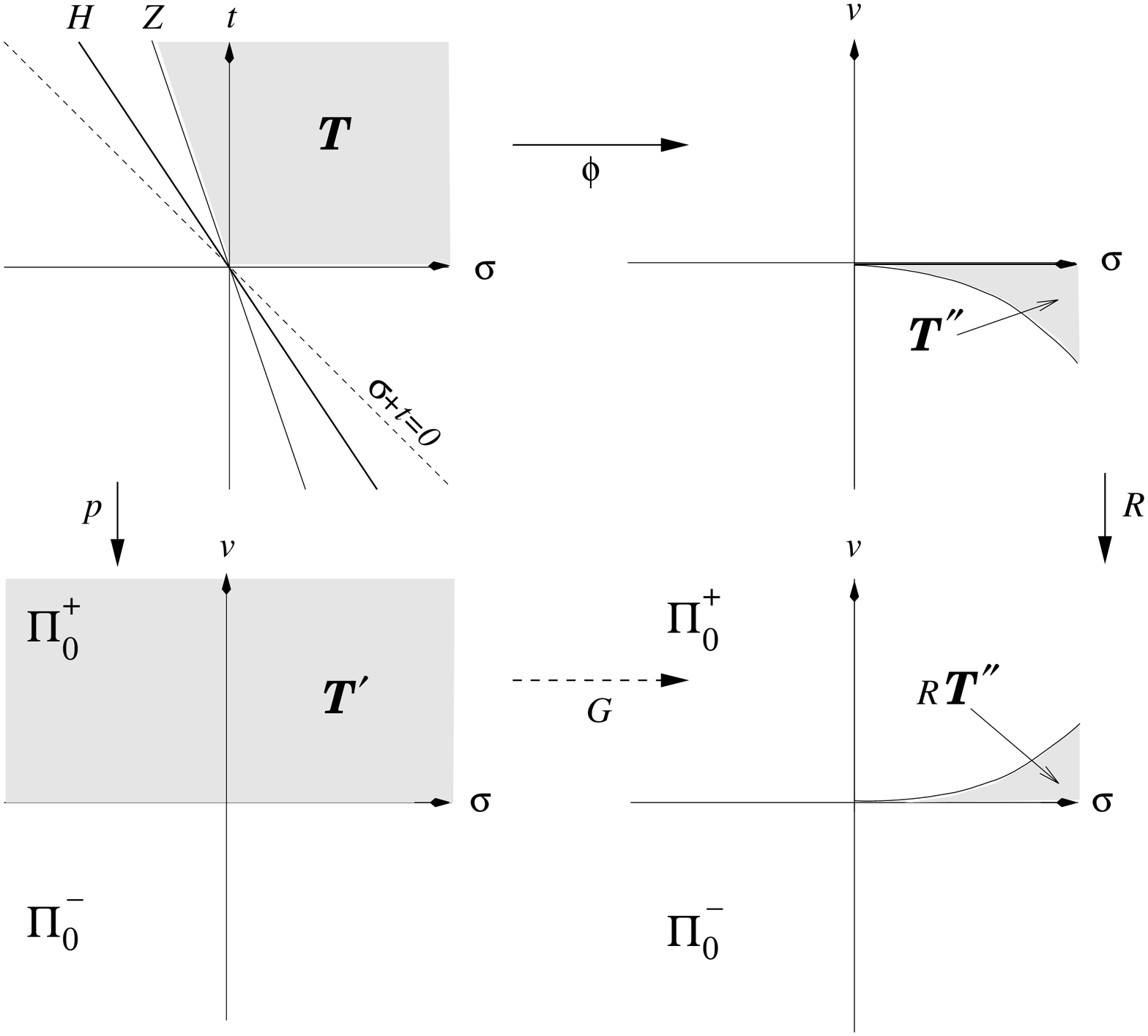}}
\end{center}
  \caption{The actions of $p$,$\phi$ and $R$ close to $(\tau_*,0)$ 
           with $a'(\tau_*)<0$: here $\tau=\tau_*+\sig$. The shaded
           area in the top left diagram represents the visible part of
           $V'$, and the other shaded areas 
           are its images under the relevant maps as indicated.} 
   \label{fi:fig6}
\end{figure}
From this geometry it is reasonable to suppose that under
iteration of $G$ all points of $\T'$ will tend to $\Pi_0\,$.
We now look more closely to verify that this is the case.
\msk

Consider an arc $\alpha$ in $V'$ given in $(\sig,t)$ coordinates by
$t=\ell\sig$ with $\ell>0$ and $\sig\ge0$.   
From~(\ref{e:pnew}) and~(\ref{e:phinew}) we have
\begin{eqnarray}
p(\sig,\ell\sig) &
   = \bigl(\sig,-k(3\ell+\ell^2)\sig^2+\ell\sig^3\mu(\ell,\sig\bigr)\\
q(\sig,\ell\sig) &
   =\bigl(\sig(1+\ell),-rk(3\ell+2\ell^2)\sig^2+\ell\sig^3\nu(\ell,\sig)\bigr)
\end{eqnarray}
where $\mu,\nu$ are continuous (in fact smooth) functions for all $\ell$
and sufficiently small $\sig$ (depending on $\ell$).
\msk

The images of $\alpha$ under $p$ and $q$ are locally the graphs of two
smooth functions $v=\tilde p(\sig)$ and $v=\tilde q(\sig)$.  Choose
$\delta>0$ and constants $K>0,L>0$ such that
\[
|\mu(\ell\sig)|\le K\,\qquad |\nu(\ell,\sig(1+\ell)^{-1}|\le L
\]
for  $0\le \ell\le -\tfrac94 kr$ and $0\le\sig\le\delta$.  We then find
that if $\delta$ is small enough so that
\begin{eqnarray}
K\delta < -k(3+\ell)\tfrac13(1-r)  \\
L\delta < -rk(3+2\ell)(1+\ell)\tfrac13(1-r)
\end{eqnarray}
we have 
\begin{eqnarray}
\til p(\sig) &> -k(3\ell+\ell^2)\tfrac13(2+r) \\
\til q(\sig) &< -rk(3\ell+2\ell^2)(1+\ell)^{-2}\tfrac13(1+2r)
\end{eqnarray}
for $0\le\sig\le\delta$ (remember that $k<0$).
Hence
\[
\til q(\sig) / \til p(\sig) < Mr(1+2r)(2+r)^{-1}
\]
where $M=(3+2\ell)(3+\ell)^{-1}(1+\ell)^{-2}$.  It is easy to check that $M<1$
for $\ell>0$ and also that $r(1+2r)<2+r$ for $0<r<1$.  Hence for
$0\le\sig<\delta$ we have
\[
\til q(\sig) < N \til p(\sig)
\]
where $0<N<1$.  Thus we obtain the expected result:
\begin{prop}
Under iteration of $G$ the arc $p(\alpha)$ tends $C^0$-uniformly to
the $\tau$-axis over the interval $[0,\delta]$.
\endproof
\end{prop}
Using the results from Section~\ref{s:nondegen} we deduce :
\begin{cor}
For $\sig\in[0,\delta]$ and $\tau=\tau_*+\sig$
the stable manifold $W_\tau^s$ extends
backwards to a point $(\tau',0)\in\Pi_0$ with $\tau'<\tau_*\,$.
\endproof
\end{cor}
Hence we have the following local picture of the dynamics of the
impact oscillator.
\begin{theo}  \label{t:foliate}
Let $w_*=(\tau_*,0)\in\Pi_0$ be a tangency point with
$a'(\tau_*)<0$. Then there is a \nhd\ $U$ of $w_*$ in $\ppb$ 
such that $U^+:=U\cap\Pi^+$ is smoothly foliated
by arcs of stable manifolds of points $(\tau,0)\in\Pi_0$ with
$\tau>\tau_*$, each arc having end-points on $\Pi_0$ on opposite sides
of $\tau_*\,$.  See Figure~\ref{fi:fig7}.
\end{theo}
Note in particular that each sufficiently small arc
$[w_1,w_*]\subset\Pi_0$ is taken by $G$ to a smooth arc $\epsilon$ tangent 
to $\Pi_0$ at $w_*$ in the region $\sig>0,v>0$. 

\begin{figure}[!ht]  
\begin{center}
  \scalebox{0.6}{\includegraphics{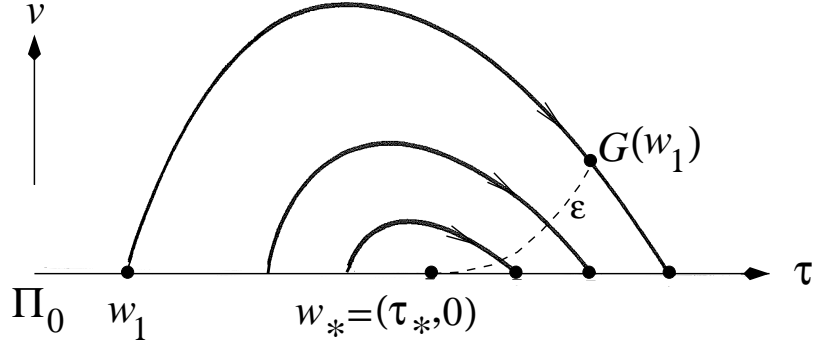}}
\end{center}
  \caption{Chatter close to $w_*=(\tau_*,0)$ with $a'(\tau_*)<0$. 
           Points of $\Pi_0$ with $\tau>\tau_*$ are limit points of
           nondegenerate chatter that becomes increasingly degenerate
           as $\tau\to\tau_*$ from above. The arc $[w_1,w_*]$ on
           $\Pi_0$ is taken by $G$ to the arc $\eps$ tangent to
           $\Pi_0$ at $w_*$.}
\label{fi:fig7}
\end{figure}
\section{Further degeneracy: dynamics near a swan point}
Finally we consider the dynamics of $G$ close to a swan point
$(\tau_*,0)\,$; here $a(\tau_*)=a'(\tau_*)=0$ but $a''(\tau_*)\ne0$.

As before, we work with coordinates
$\tau=\tau_*+\sig$ and we write $(\sig,\eta)=(s,sm)=\chi(s,m)$.  
Using~(\ref{e:swan}) we see that 
the maps $p$ and $\phi$ then become
\begin{eqnarray}
   p:(s,m)&\mapsto\bigl(s,\ell s^3(6m+4m^2+m^3+O(s))\bigr)
   \label{e:pdeg3} \\
\phi:(s,m)&\mapsto\bigl(s(1+m),-\ell s^3(6m+8m^2+3m^3+O(s))\bigr)
\end{eqnarray}
where $\ell=-\tfrac1{24}a''(\tau_*)$ and where the terms $O(s)$
denote smooth functions of $(s,m)$ that vanish when $s=0$. 

In $(s,m)$ coordinates the map $h$ becomes 
$\bar h:(s,m)\mapsto(\bar s,\bar m)$ where
\begin{eqnarray}
  s(1+m) &= \bar s  \\
 rs^3(6m+8m^2+3m^3+O(s))&=\bar s^3(6\bar m+4\bar m^2+\bar m^3+O(s))
\end{eqnarray} 
and substituting the first equation into the second for $s\ne0$ we obtain
\begin{equation*}
 r(6m+8m^2+3m^3)=(1+m)^3(6\bar m+4\bar m^2+\bar m^3)+ O(s).
\end{equation*}

The graphs of the two real-valued functions
\begin{eqnarray}
f:m&\mapsto r(6m+8m^2+3m^3)(1+m)^{-3} \\
g:m&\mapsto 6m+4m^2+m^3
\end{eqnarray}
intersect at a unique point $m_*$ which lies in $(-1,0)$.
This proves the following.
\begin{prop}
The map $\bar h$ has a fixed point at $(0,m_*)$ and no other fixed point on
the $m$-axis.
\end{prop}
It is easy to check from the graphs of $f,g$ that $f'(m_*)>g'(m_*)$,
and also to verify that
\begin{equation}
    D\bar h(0,m_*)=  \begin{array}{cc}
      1+m & 0 \\    * &  f'(m_*)/g'(m_*)
                 \end{array}.
\end{equation}
This implies the next result.
\begin{prop}
The map $\bar h$ has a hyperbolic saddle fixed point at $(0,m_*)$. 
The $m$-axis is
the unstable manifold, while the stable manifold is a smooth curve $W^s$
through $(0,m_*)$ transverse to the $m$-axis.
\end{prop}
The interpretation of these results now depends on the sign of
$a''(\tau_*)$.
\medskip  

If $a''(\tau_*)<0$ then $a(\tau)$ has a local maximum at $\tau_*$ with
$a(\tau_*)=0$, and so from~{\ref{e:acc}) we have $\dbd{}t\,v_c(\tau,t)>0$ 
for all $(\tau,t)$ in some \nhd\ of $w_*=(\tau_*,0)$ in $\Pi$ except
at $w_*\,$. Therefore $F$ is defined on a
\nhd\ of $w_*$ in $\Pi$ and is a smooth local homeomorphism onto a
\nhd\ of $w_*$ in $V'$.  Hence if 
$\beta=\chi(W^s)$ then $\gamma=p(\beta)\cap\ppb$ is a smooth
invariant curve $\gamma$ for $G$ terminating at the swan point.  As is clear
from~(\ref{e:pdeg3}) this curve $\gamma$ has cubic tangency with
$\Pi_0$ at $w_*$.
\begin{theo}  \label{t:t3}
Let $w_*=(\tau_*,0)\in\Pi_0$ be a swan point with $a''(\tau_*)<0$. The
map $G$ has a unique invarant arc $\gamma$ with end-point at $w_*\,$, the arc
being smooth and cubically tangent to the $\tau$-axis at $w_*$ from
the side $\tau<\tau_*$ and $v>0$.  The fixed point $w_*$ is
exponentially attracting for $G|\gamma$, and is weakly normally hyperbolic
(repelling) for $G$. All points in a \nhd\ of $w_*$ but not on
$\gamma$ lie on stable manifolds of regular points of $\Pi_0$
(nondegenerate chatter).  See Figure~\ref{fi:fig8}.
\endproof
\end{theo} 
Thus with $a''(\tau_*)<0$ all initial data close to $w_*$ lead to nondegenerate
chatter, with the exception of those on the curve $\gamma$ which
exhibit a form of highly degenerate chatter.
\medskip

If on the other hand $a''(\tau_*)>0$ then we have $v_c(\tau,t)<0$ for
sufficienatly small $t>0$ and $\tau$ close to $\tau_*$, and so the
image under $F$ of a \nhd\ of $w_*$ in $\ppb$ will include no points
close to $w_*$. Therefore the local analysis of low-velocity impacts
which is the subject of this paper will not apply.
\begin{figure}[ht!]  
\begin{center}  
\scalebox{0.6}{\includegraphics{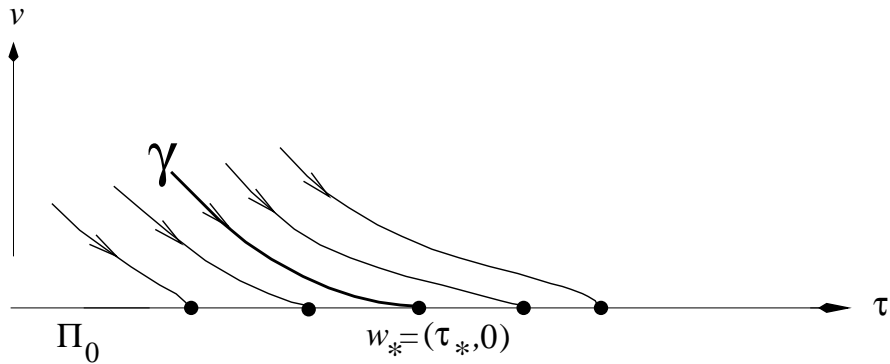}}
\end{center}
  \caption{Stable manifolds of chatter points close to a swan point
  $(\tau_*,0)$ where $a'(\tau_*)=0$ and $a''(\tau_*)<0$.}  
\label{fi:fig8}
\end{figure}
\section{Unfolding the dynamics at the swan point}
Typically (the precise conditions on the functions $f$ and $g$
in~(\ref{e:eqn}) are described in~\cite{DC}) swan points will occur for
only certain isolated values of $c$ or of some other parameter in the
system: swan points are codimension-$1$ phenomena.  Therefore it is
important to consider the bifurcations in the local dynamics 
as a generic transition through a swan point takes place.
\medskip

We shall take the parameter to be $c$ for simplicity, although this is not
crucial for the discussion.  For clarity we also now re-introduce the $c$
suffix in notation for the maps $p_c,\phi_c,F_c$ and $G_c$ and the
surface $V_c'$. We suppose that
$a_c(\tau_*)=a_c'(\tau_*)=0$ when $c=c_0$ and let
\[
d(c,\tau):=a_c''(\tau)\,\dbd ac(c,\tau)
\]
where $a(c,\tau)=a_c(\tau)$.
As $c$ increases through $c_0$ the function $a_c(\tau)$ undergoes a
Morse transition at which a pair of zeros $\tau_1,\tau_2$ of $a_c(\tau)$
is created or annihilated according to the sign of
$d_0:=d(c_0,\tau_*)$.
Note that under the Correspondence Principle the creation of two
tangency points at a swan point corresponds to the creation of two
cusps at a swallowtail (Figure~\ref{fi:fig2}(c)).
\msk

Without loss of generality we suppose $d_0<0$.  Thus
as $c$ increases through $c_0$ an interval $(\tau_1,\tau_2)$ is
created on which $a_c(\tau)$ reverses its sign, and a new component of
$P_c$ (a {\em swan} configuration) is created out of the point
$w_*\,$ as described in~\cite{DC}.  See Figure~\ref{fi:fig9}.
The creation of the swan is in fact a version of the lips transition
(Figure~\ref{fi:fig2}(a)) with particular geometry dictated by the
context, a characteristic geometry of singular sets noted earlier by
Bruce~\cite{BR89}.
\msk

There are two cases to consider, according to whether 
$a_0''(\tau_*):=a_{c_0}(\tau_*)>0$ ({\em positive} swan point) or 
$a_0''(\tau_*)<0$ ({\em negative} swan point).  
\begin{figure}[ht]  
\begin{center}
\scalebox{0.6}{\includegraphics{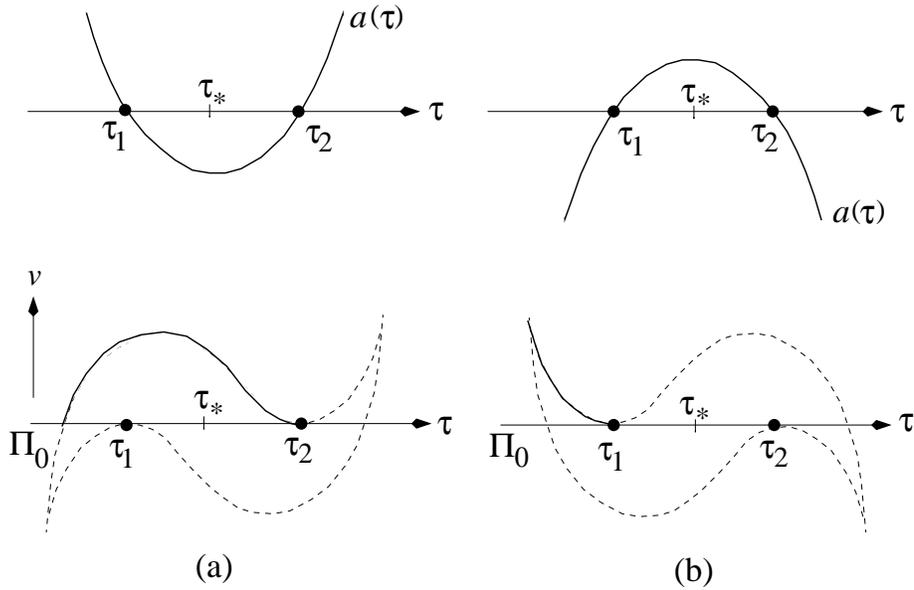}}
\end{center}
\caption{Graphs of $a(\tau)$ and associated swan configurations in
    the plane $\Pi$ in two cases: (a) positive, (b) negative.}
\label{fi:fig9}
\end{figure}
\medskip

\begin{figure}[ht]  
\begin{center}
 \scalebox{0.58}{\includegraphics{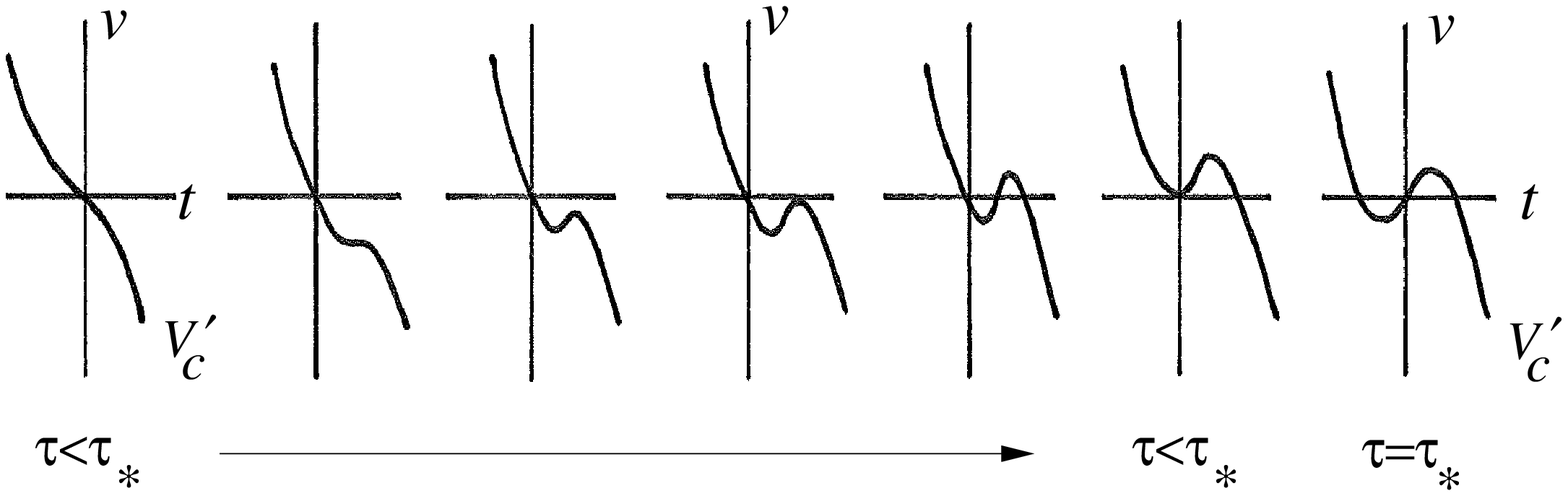}}
\end{center}
  \caption{Sections through $V_c'$ in the $(t,v)$-plane for increasing
  values of $\tau$ with $\tau\le\tau_*$.}  
 \label{fi:fig10}
\end{figure}
\begin{figure}[ht!]  
\begin{center}
  \scalebox{0.48}{\includegraphics{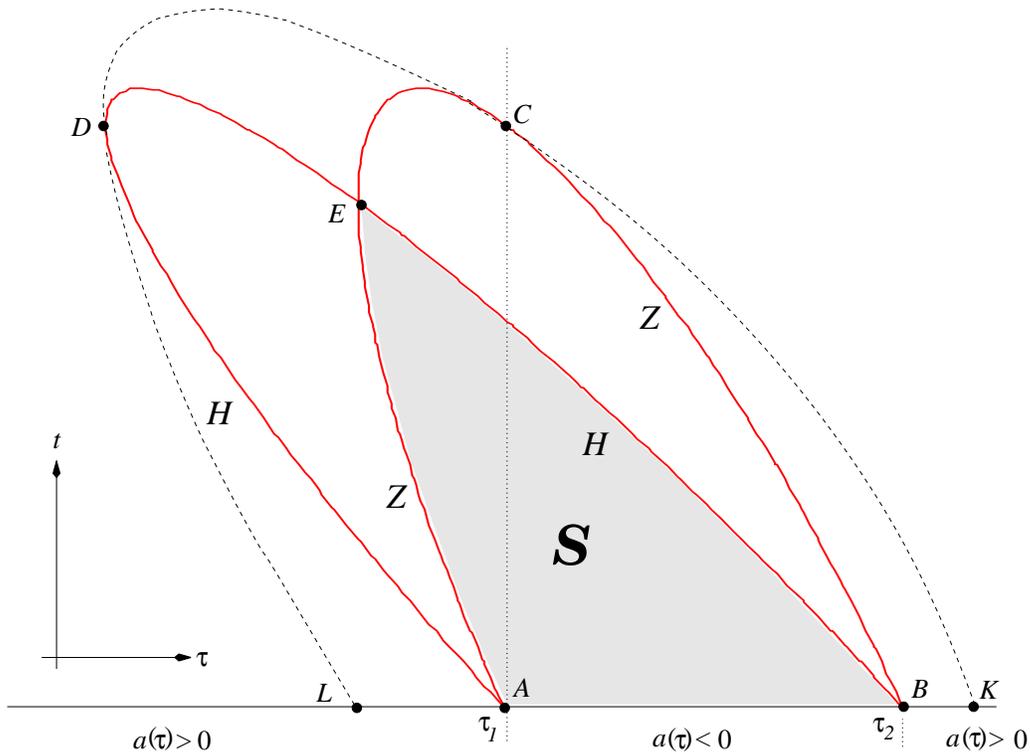}}
\end{center}
  \caption{Positive swan configuration: the visible part $S$ of $V'$
  is bounded by an arc of the horizon $H$ and an arc of the zero locus $Z$.
  See text for details.}  \label{fi:fig11}
\end{figure}
\subsection{Positive swan point: $a''(\tau_*)>0$}
For $c<c_0$ there are no low-velocity impacts occuring close to $w_*$,
and for $c=c_0$ although the acceleration vanishes at $\tau_*$ there
are still no low-velocity impacts.  We therefore focus on the
low-velocity dynamical phenomena that are created when $c>c_0\,$. 
Once again we now suppose such $c$ fixed and dispense with it as a suffix.

\msk

To help picture the dynamics of $G$ (and in particular the geometry
of $F$) we show in Figure~\ref{fi:fig10} a sequence of sections through $V'$ at
constant $\tau$ for various choices of $\tau\le\tau_*$.  This
information can be obtained from the Taylor expansion~(\ref{e:vtaylor}),
regarding $v_c(\tau_*+\sig,t)$ as a function of $t$ unfolded by the
parameter $\sig$.  Analogous sketches for $\tau\ge\tau_*$ are obtained
by reversing the signs of both $t$ and $v$. 
\msk

In Figure~\ref{fi:fig11} we show the horizon $H$ and the zero set $Z$
in $(\tau,t)$-coordinates on $V'\,$, thus viewed from the \lq
vertical' $v$ direction. 
Here $A=(\tau_1,0)$ and $B=(\tau_2,0)$ where $\tau_1<\tau_2$ are the
two zeros of $a(\tau)$ near $\tau_*$ for $c>c_0$ and close to $c_0$.
The shaded area $\S$ indicates the connected component of the 
visible part of $V'$ that contains the interval $AB$ (see Section~\ref{ss:hit}).
\msk

As already seen in Chapter~\ref{s:degen}, the dynamics of $G$ close
to $A$ and $B$ are as in Figure~\ref{fi:fig5} and Figure~\ref{fi:fig7}
respectively.  These combine to give the swan dynamics in the way that
we now describe. 
\msk

Figure~\ref{fi:fig12}(a) shows the same configuration as
Figure~\ref{fi:fig11} but viewed along the $t$-axis, or in other words 
its image in $\Pi$ under the projection $p\,$.  
Points $p(C),p(D),\ldots$ are denoted $C',D',\ldots\,$; 
note that $A'=A$ and $B'=B$.  The arc $BE$ of $H$
is taken by $p$ to a smooth arc $BE'$ in $\ppb$ tangent to $\Pi_0$ at $B$ and
transverse to $\Pi_0$ at $E'$:  together with the interval $E'B$ of
$\Pi_0$ it bounds the simply-connected region $\S'=p(\S)\subset\ppb$.
\msk

Figure~\ref{fi:fig12}(b) shows the image of the configuration in
Figure~\ref{fi:fig11} under the re-set map $\phi$, with points
$\phi(C),\phi(D),\ldots$ denoted $C'',D'',\ldots\,$.  The arc $AE$ of
$Z$ is taken by $\phi$ to a smooth arc $AE''$ in $\pnb$ tangent to $\Pi_0$ at
$A$ and transverse to $\Pi_0$ at $E''$ (part of the apparent outline
in view of the Correspondence Principle: see
Proposition~\ref{p:corresp}); together with the interval $AE''$ of
$\Pi_0$ it bounds the simply-connected region
$\S''=\phi(\S)\subset\pnb\,$.
\msk

To construct the dynamics of $G$ we compose the {\it inverse} of $p$ with
$\phi$ and then with the restitution $R$.
The latter takes the arc $AE''$ in $\pnb$ to a smooth arc in
$\ppb$ tangent to $\Pi_0$ at $A$ and transverse to $\Pi_0$ at $E''$.
Its intersection with $\S'$ consists of a smooth arc $\varepsilon$
also tangent to $\Pi_0$ at $A$ and with its other end-point $N$ lying
on the arc $LB$ of the apparent outline.

The $G$-invariant curve $\gamma$ at the degenerate chatter point $B$
is tangent to $\Pi_0$ at $B$ and transverse to $\Pi_0$ at its other point of
intersection. It separates $\S'$ into two regions:
one foliated by stable manifolds of points of $AB$ (nondegenerate
chatter), the other consisting of points that eventually leave $\S'$.  
See Figure~\ref{fi:fig13}. 
\begin{figure}[ht!]  
\begin{center}
\scalebox{0.6}{\includegraphics{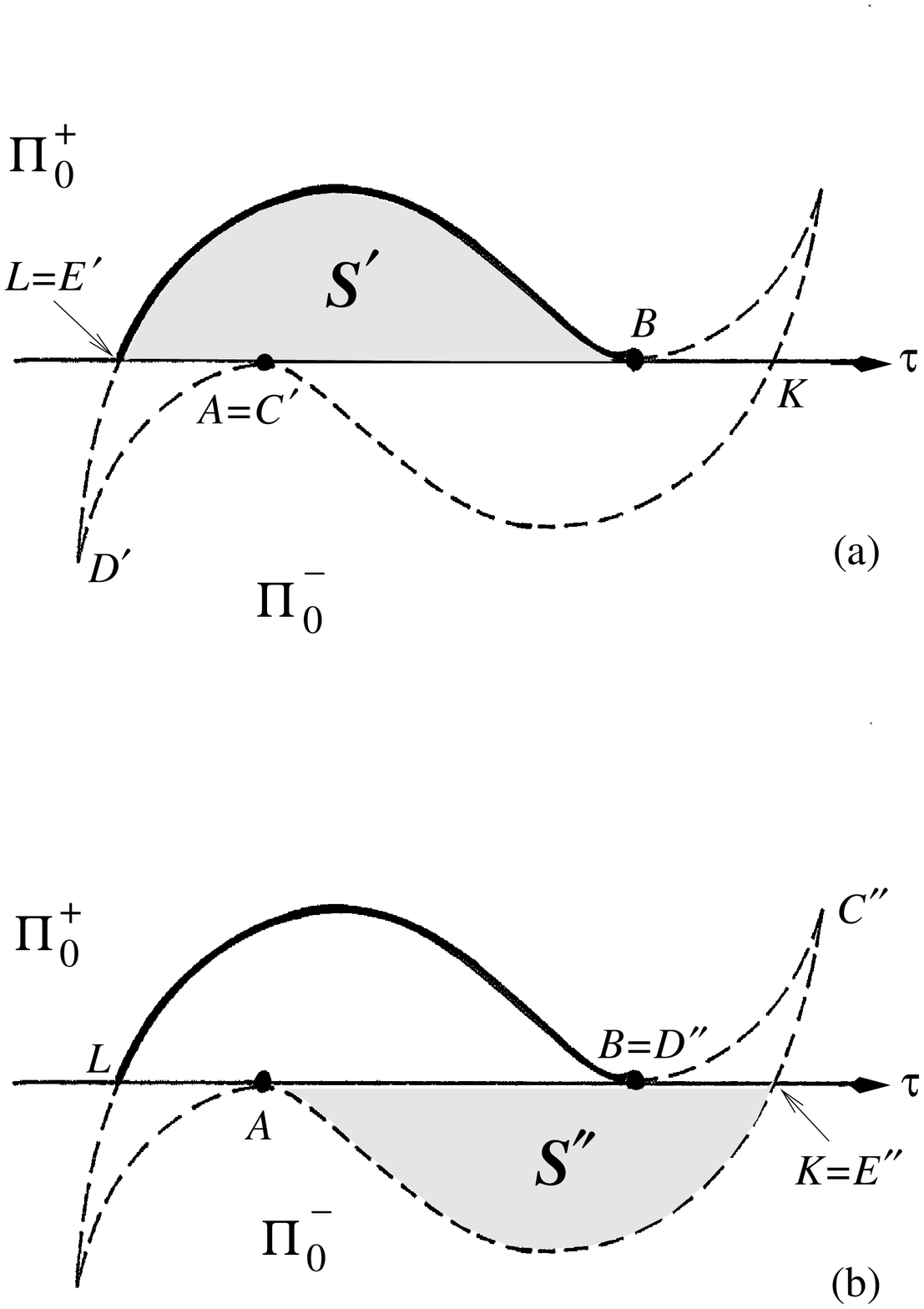}}
\end{center}
  \caption{The images under (a) the projection $p$ and (b) the re-set
    map $\phi$ of various regions of $V'$ in Figure~\ref{fi:fig11} for
    an unfolded positive swan point: see text for details.}  
\label{fi:fig12}
\end{figure}
\begin{figure}[ht!]  
\begin{center}
  \scalebox{0.6}{\includegraphics{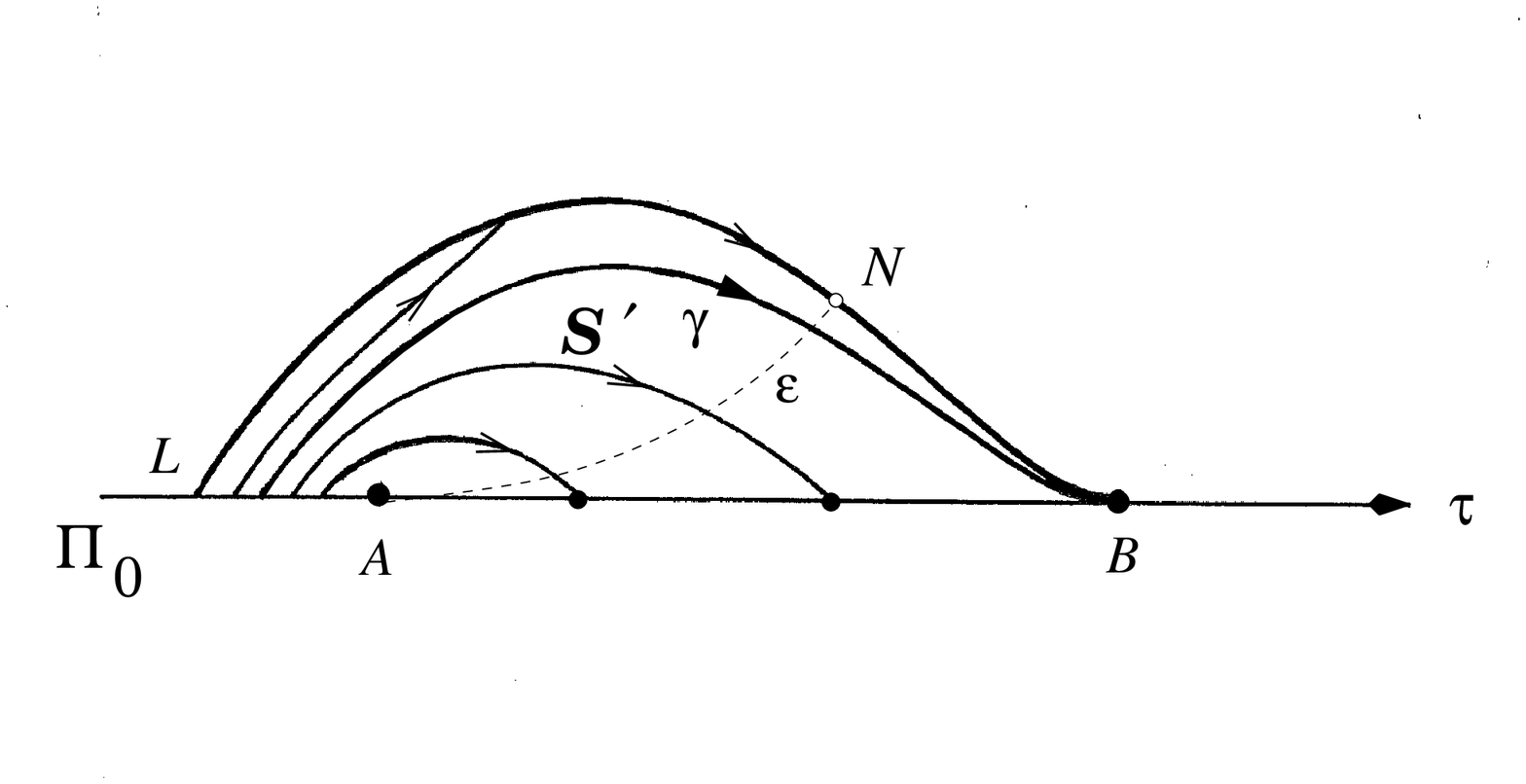}}
\end{center}
  \caption{Invariant foliation for dynamics of $G$ in the region $S'$.
  Outside $S'$ the unfolding of the swan point gives no local information.}  
\label{fi:fig13}
\end{figure}
\subsection{Negative swan point: $a_0''(\tau_*)<0$}
For $c<c_0$ we have $a(\tau)<0$ for all $\tau$ in a \nhd\ of $\tau_*$ and so
Theorem~\ref{t:t1} applies: every point lies in the stable manifold of
some point of $\Pi_0$ and nondegenerate chatter takes place locally
as in Figure~\ref{fi:fig3}(b).

For $c=c_0$ Theorem~\ref{t:t3} applies: there is a smooth stable manifold for
$(\tau_*,0)$ that is cubically tangent to $\Pi_0$ and corresponds to
highly degenerate chatter, while all other orbits experience
nondegenerate chatter becoming increasingly degenerate as
$\tau\to\tau_*$ from either side as in Figure~\ref{fi:fig8}.  
We therefore focus on the case $c>c_0\,$.
\msk

The analogue of Figure~\ref{fi:fig10} for this case is obtained by
inverting the $v$-axis.  In Figure~\ref{fi:fig14} we again show the
horizon $H$ and the zero set $Z$ as viewed from the $v$-direction, 
with $A,B$ as before corresponding to
the values $\tau_1,\tau_2$ of $\tau$ where $a(\tau)=0$. Now, however,
it is a different region $\T=\T_1\cup\T_2\cup\T_3$ that is visible, 
bounded by the arc $BC$ of $Z\,$, the arc $DA$ of $H\,$, and the
(dotted) arc $DC$ which is the {\em shadow} of $DA$, 
that is the set of ``next-hit'' points starting
from $DA$: more precisely it is the set of points $(\tau,v;t_1)\in V'$ 
for which there exists $t_0>0$ with $(\tau,v;t_0)\in DA$ but
$(\tau,v;t)\notin V'$ for all $t_0<t<t_1$. 
The visible part of $V'$ is the region lying
outside the loop $ADCBA$ (including $AD$ but not including the open arc $DC$).
\msk

\begin{figure}[ht!]  
\begin{center}
\scalebox{0.48}{\includegraphics{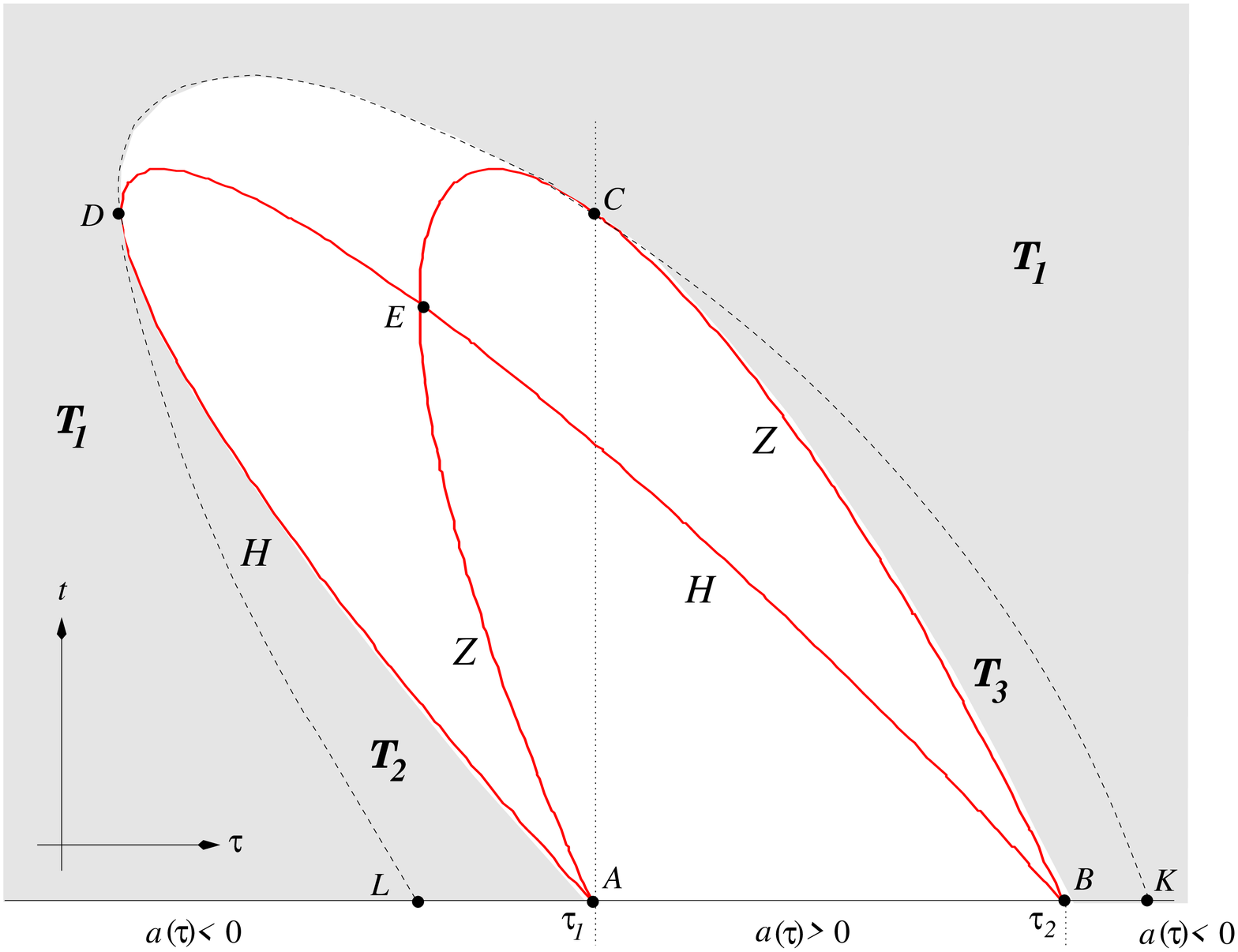}}
\end{center}
  \caption{Negative swan configuration on $V'$: the visible part
    (shaded) of $V'$ here is bounded by an arc of $H$
    and an arc of $Z$ together with an arc $DC$ of the shadow of $H$.}
\label{fi:fig14}
\end{figure}

Under the correspondence principle in Figure~\ref{fi:fig14} 
the points $A,B$ each correspond to cusp
points elsewhere on $H$: in fact $B$ corresponds to $D$ 
while $A$ corresponds to its counterpart (not shown) with
$t<0$. Figure~\ref{fi:fig15}(a) shows the images $A',\ldots,E'$ of
the points $A,\ldots,E$ of $V'$ under projection $p$, while
Figure~\ref{fi:fig15}(b)shows their images $A'',\ldots,E''$ under the
re-set map $\phi\,$.
\begin{figure}[ht!]  
\begin{center}
\scalebox{0.6}{\includegraphics{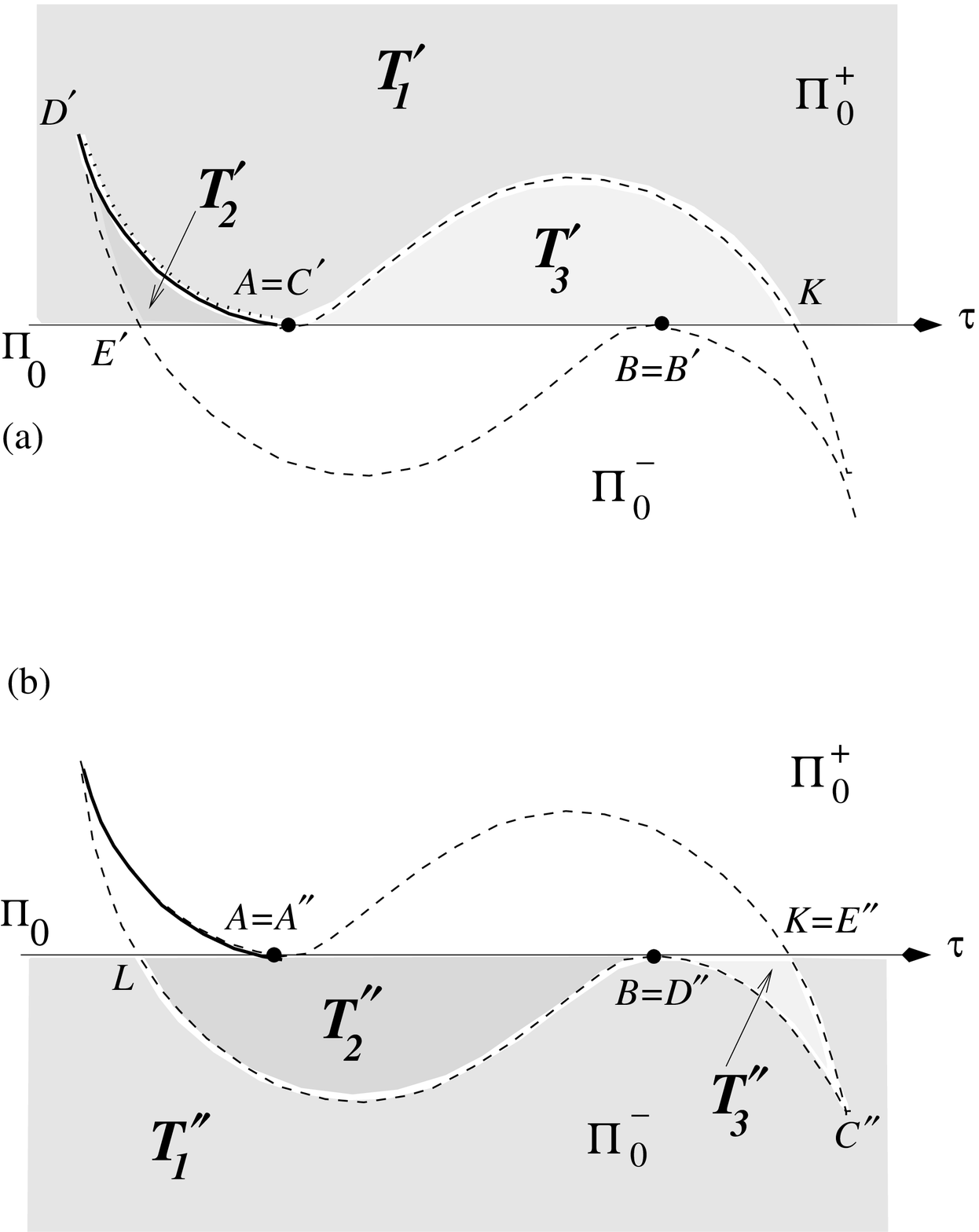}}
\end{center}
\caption{The images under (a) the projection $p$ and (b) the re-set
    map $\phi$ of various regions of $V'$ for an unfolded negative
    swan point: see text for details.}  
\label{fi:fig15}
\end{figure}
The only visible part of $H$ is the
arc from $A$ to $D$: under $p$ this projects to the arc $AD'$ in
$\ppb$ while under $\phi$ it is taken to the interval $A''D''$ on
the $\tau$-axis $\Pi_0$.
\msk

At points of the shadow arc $CD$ apart from $D$ the velocity $\dot x_c$ is nonzero 
(in fact it has to be negative) and so $\phi$ takes $CD$ 
to an arc $C''D''$ with $v<0$ except at $D''$; 
moreover $C''$ is a cusp point of $P$ by Proposition~\ref{p:phising}.
From this we see the nature of the
discontinutity of the map $\phi\circ F$ (and hence of $G\,$):
cutting $\Pi^+$ along the arc $AD'$ then mapping one edge (heavy) to
the interval $A''D''$ of $\Pi_0$ and the other edge (dotted) to the arc
$D''C''$.  Meanwhile the interval $AB$ of Figure~\ref{fi:fig15}(a)
is mapped by $F$ to the arc $CB$ of $Z$ which is then taken by $\phi$
to the arc $C''B$ of Figure~\ref{fi:fig15}(b).
\msk

These pictures repay contemplation in order to form a clear mental 
picture of the dynamics of $G\,$. Figure~\ref{fi:fig16} shows schematically
how $\phi\circ F$ takes each segment of the swan region into the
next segment; remember, however, that the map $G$ also incorporates the
restitution $R$ at each step and takes each of $\ppb,\pnb$ into
itself.  We now describe the dynamics.
\medskip

\begin{figure}[ht]  
\begin{center}
\scalebox{0.6}{\includegraphics{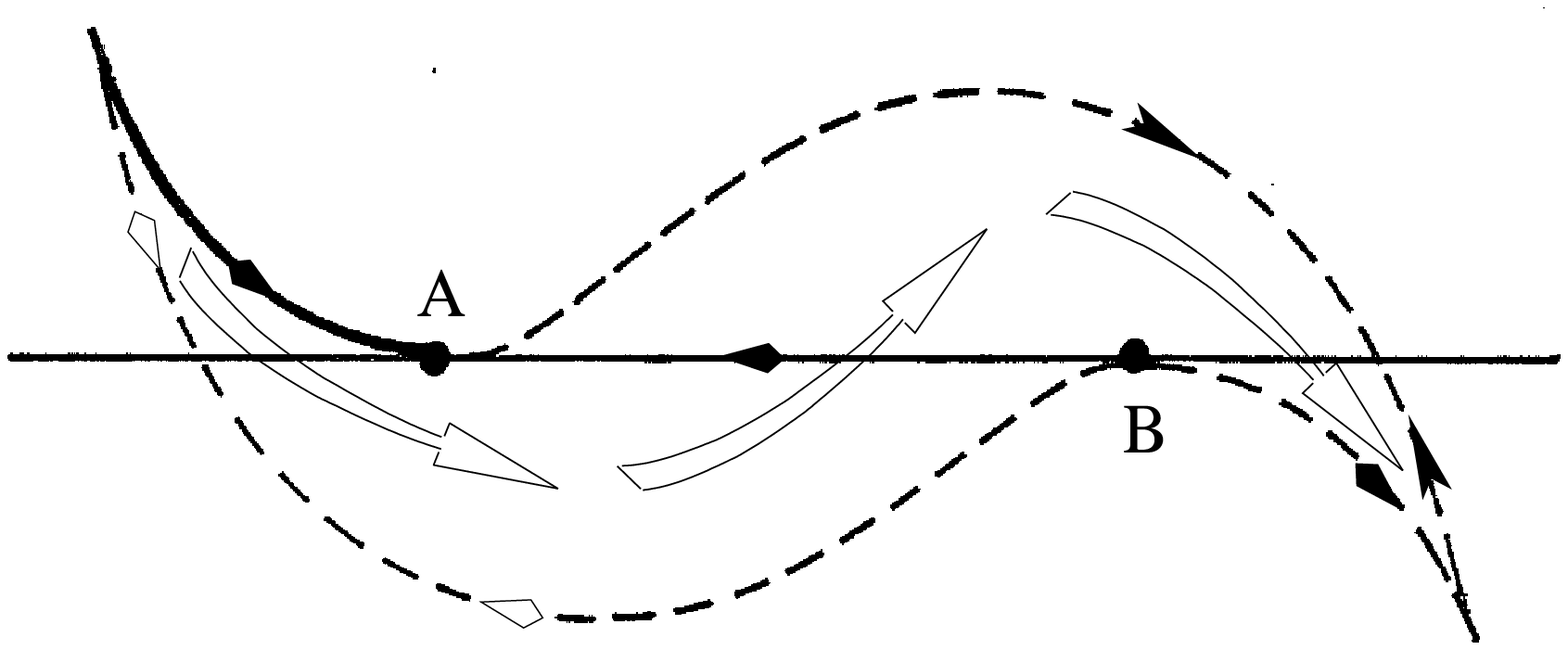}}
\end{center}
  \caption{Schematic representation of the action of $\phi\circ F$
  within the swan region. Matching arrows indicate arcs and their
  images.    The third step is discontinuous along $AB$.}  
\label{fi:fig16}
\end{figure}

In Section~\ref{ss:tang} we have already obtained local models for the 
dynamics of $G$ close to the two tangency points $A$ and $B$
(Figure~\ref{fi:fig5} and Figure~\ref{fi:fig7} respectively).
As there are no discontinuities of $G$ close to $B$, the \nhd\ of
$B$ that is foliated by stable manifolds $W_\tau^s$ of points $(\tau,0)$ with
$\tau>\tau_2$ (by Theorem~\ref{t:foliate}) must extend at least as far as the
stable manifold $W_{\tau_3}^s=W^s(X)$ that contains the point $A$, where
$X=(\tau_3,0)$ with $\tau_3>\tau_2$.  See Figure~\ref{fi:fig17}.
\begin{prop}
The stable manifold $W^s(X)$ is tangent to $\Pi_0$ at $A$.
\end{prop}
\proof
We argue that $F\bigl(W^s(X)\bigr)\subset V'$ is tangent to $Z$ at $C$,
from which the result follows by applying $p\,$.

Since the map $\phi$ has a cusp singularity at $C$ by 
Proposition~{\ref{p:phising},
if $F\bigl(W^s(X)\bigr)$ were not tangent to $Z$ at $C$ then
$\phi\circ F\bigl(W^s(X)\bigr)$ would be aligned with the
direction of the cusp at $C''$ in Figure~\ref{fi:fig15}.  In that case 
$R\circ\phi\circ F\bigl(W^s(X)\bigr)=G\bigl(W^s(X)\bigr)=W^s(X)$ 
would be tangent to the arc $\varepsilon'=R(D''C'')$ at its endpoint $R(C'')$,
and this is not consistent with the geometry of Figure~\ref{fi:fig7}:
stable manifolds of points of $\Pi_0$ to the right of and close to $B$
are transverse to $\varepsilon'$.

\endproof
\medskip

Let $\lam$ denote the arc of $W^s(X)$ from $A$ to $Q=W^s(X)\cap\varepsilon'$.  
The inverse image $G^{-1}(\lam)$ is a closed loop around the
discontinuity arc $\delta=AD'$ and with both ends at $A$.
The inverse images of $\delta$ under iterates of
$G$ form a sequence of arcs tangent to $\Pi_0$ at the degenerate
chatter point $A$ and accumulating on its invariant arc $\gamma$,
and therefore the inverse images 
of $\lam$ form a sequence of loops also accumulating on $\gamma$.   

Stable manifolds $W^s(Y)$ of points $Y$ to the right of $X$ on $\Pi_0$
(i.e. with $\tau>\tau_3$)
experience no discontinuities locally: those for $Y$ close to $X$ 
oscillate more and more wildly under backwards iteration of $G$ as they
attempt to adhere to the loops of $G^{-n}(\lam)$.  See Figure~\ref{fi:fig17}.
\begin{figure}[ht!]  
\begin{center}
\scalebox{0.6}{\includegraphics{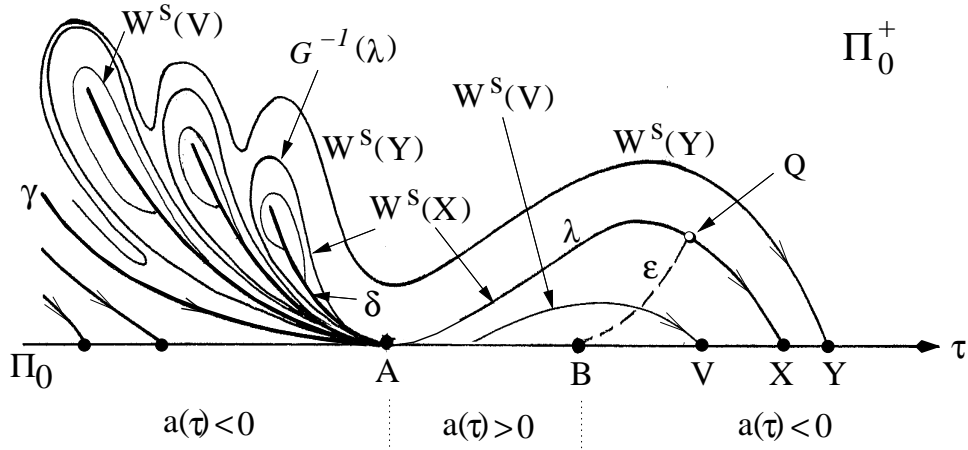}}
\end{center}
\caption{Dynamics of $G$ created at unfolding of a negative swan
  point: see text for details.}  
\label{fi:fig17}
\end{figure}
\section{Global dynamics}
The dynamical phenomena that become apparent in the unfolding of a
negative swan point give some clues about the global geometry for
dynamics of an impact oscillator.  In particular we see that
the particular geometry of the smooth map $p\circ\phi^{-1}$
close to a tangency point $w_*=(\tau_*,0)\in\Pi_0$ (so that 
$z=\phi^{-1}(w_*)$ is a cusp point of $H$) plays a key role.  
At a tangency point $w_*$ where $a'(\tau_*)>0$ the nearby stable
manifolds $W^s_\tau$ with $\tau>\tau_*$ have infinitely many loops
accumulating on the invariant curve $\gamma$, that is the stable
manifold of $w_*\,$  Therefore we see that the global behaviour of the 
stable manifolds of degenerate chatter points,
and in particular their intersections with unstable manifolds of fixed
or periodic points elsewhere in the phase space $\Pi_0^+\,$,
are crucial in organising the global effects of the complicated dynamics
unleashed at a (negative) swan bifurcation.  The swan
configuration created at a negative swan point will persist until
variation in the parameter $c$ causes it to impinge on some other
branch of the apparent outline $P\,$.  
Careful study of the generic ways in which this
can happen will provide important tools for understanding the
complicated global structure of discontinuous planar discrete dynamical
systems modelling $1$-degree of freedom impact oscillators.  
\bsk

\ack
I am most grateful to the Centre de Recerca Matem\`atica, Barcelona, for
hospitality during the preparation of the main part of this paper as
part of the Research Programme on Complex Non-Smooth Dynamical Systems
organised there by the Bristol Centre for Applied Nonlinear Mathematics in
Spring 2007. I thank colleagues at BCANM for hospitality 
and continuing interest in and support for this research.


\end{document}